\documentclass[12pt]{iopart} 
\usepackage{iopams}
\newcommand{\ket}[1]{\left|#1\right\rangle}
\newcommand{\bra}[1]{\left\langle#1\right|}
\newcommand{\ii}{\mathrm{i}}
\newcommand{\dd}{\mathrm{d}}
\newcommand{\eqref}[1]{(\ref{#1})\,}

\begin{document}
\title[Quantum quenches in the sinh-Gordon and Lieb--Liniger models]{Quantum quenches in the sinh-Gordon and Lieb--Liniger models}
\author{Emanuele Di Salvo and Dirk Schuricht}
\address{Institute for Theoretical Physics, Center for Extreme Matter and Emergent Phenomena, Utrecht University, Princetonplein 5, 3584 CE Utrecht, The Netherlands}
\ead{e.disalvo@uu.nl, d.schuricht@uu.nl}

\begin{abstract}
The non-relativistic limit of integrable field theories at equilibrium has been intensively studied in the previous years; the simplest non-trivial case relates the sinh-Gordon model to the Lieb--Liniger model. Here we study this non-relativistic limit out of equilibrium, namely in the time evolution after a quantum quench. The obtained results agree with the known ones for the Lieb--Liniger model, thus showing that the non-relativistic limit is applicable in this out-of-equilibrium setting.
\end{abstract}
\vspace{2pc}
\noindent{\it Keywords}: quantum quench, sinh-Gordon model, Lieb--Liniger model, non-relativistic limit

%\submitto{\JSTAT}
\maketitle

%%%%%%%%%%%%%%%%%%%%%%%%%%%%%%%%%
\section{Introduction}\label{Introduction}
%%%%%%%%%%%%%%%%%%%%%%%%%%%%%%%%%
Lorentz invariance is one of the most fundamental symmetries in physics. However, in many cases the velocity of the involved particles is much smaller than the velocity of light, making the description of the system in terms of a non-relativistic theory feasible. The most prominent example for such a situation is the derivation of the Pauli equation, describing non-relativistic electrons in an external electromagnetic field, from the fully Lorentz invariant Dirac equation. 

Of particular theoretical interest are one-dimensional models~\cite{Giamarchi04}, which generically show fascinating many-particle effects. Furthermore, various numerical and analytic tools for their study are available, including field theoretical methods based on integrability. Such integrable field theories possess an infinite set of conserved quantities that constrain scattering amongst its particles to purely elastic processes~\cite{Mussardo10}. The non-relativistic limit of several of these theories has been studied in the past~\cite{Grosse-04,Calabrese-14,Bastianello-16,Bastianello-17}. Of particular interest for our purpose is the non-relativistc limit~\cite{Kormos-09,Kormos-10} of the sinh-Gordon model (ShGM), which leads to the repulsive regime of the Lieb--Liniger model (LLM). Both models describe strongly interacting $(1+1)$-dimensional quantum systems. 

From a more technical point of view, the two models are solved using different techniques: the LLM is tackled by means of the Bethe ansatz formalism~\cite{KorepinBogoliubovIzergin93}, while the ShGM can be described in terms of the bootstrap approach~\cite{Smirnov92book}. Both methods allow the calculation of matrix elements and correlation functions (and other things). The non-relativistic limit of the ShGM consists of a double limit of large velocity and small coupling, such that the relevant energy scale is kept constant. For example, starting from the form factors~\cite{KoubekMussardo93} in the ShGM, the matrix elements of the corresponding operators in the LLM are obtained.  Comparing the two methods, the bootstrap approach is more efficient in terms of computational demands; in a sense the formal neatness of the Bethe ansatz is replaced with some physical intuition and the taking of the double limit. It has to be stressed that the two methods yield the same results, even though one can generate more complicated expressions in an easier way with the bootstrap one.

Up to now there has been little progress on the understanding of the non-relativistic limit in out-of-equilibrium settings~\cite{Bertini-16jsm}. We aim to extend this by considering a quantum quench in the ShGM, i.e., the time evolution of the system starting from a prepared initial state. We then take the non-relativistic limit of the calculated one-point functions. The reasons for our choice are manifold: First, quantum quenches provide a simple out-of-equilibrium setup and can be realised in ultracold atomic gas experiments. Second, given the great interest in quenches in the LLM, there are also many results~\cite{Kormos-13pra,DeNardisCaux14,DeNardis-14,DeNardis-15,GranetEssler21} to compare to ours. Third, by this comparison one can identify relativistic and non-relativistic features of out-of-equilibrium processes such as relaxation and steady state properties. Finally, compared to other quantum field theories, the dynamics of the one-point functions in the ShGM after a global quench has not been described yet.

As mentioned before, quantum quenches in LLM are interesting for different reasons: the best-known and most pressing one is the cold atom experimental realisation of the model (for a review, see e.g., Reference~\cite{Polkovnikov-11}). Furthermore, for theorists, the system itself is the ideal playground for the study of quantum quenches, given its striking richness and the aforementioned integrability. Many questions have already been answered, but still the time evolution and its dependence on the initial state are not fully understood. For example, steady state counting statistics has not been derived yet, in contrast to the equilibrium case~\cite{Bastianello-18,BastianelloPiroli18}. One result of our paper is describing how to derive all of these quantities under integrability conditions~\cite{Piroli-17npb} by means of the non-relativistic limit.

The paper is organised as follows: in Section~\ref{sec:models} we introduce the two models we are concerned with in the rest of the paper and their relation via the non-relativistic limit. In Section~\ref{Quenching procedure and stationary state} the quench setup for ShGM is described and final and initial states are investigated; the connection between the quench problem and the boundary one is introduced and clarified in terms of thermodynamic Bethe ansatz (TBA) equations. In Section~\ref{Post-quench time evolution chap} we focus on computing the time evolution of the one-point function for different operators by using two complementary approaches, namely the linked cluster expansion and the quench action method. In Section~\ref{Lieb-Liniger quenches as non-relativistic Sinh-Gordon ones} the link between the two models in the global quench setup is discussed and observables and time evolution are extracted and compared with results from the literature. Details of the computations, methods and notations are presented in the appendices.

%%%%%%%%%%%%%%%%%%%%%%%%%%%%%%%%%
\section{The models}\label{sec:models}
%%%%%%%%%%%%%%%%%%%%%%%%%%%%%%%%%
In this chapter we introduce the two systems considered in this article, the ShGM and LLM, as well as the link between them. 

%%%%%%%%%%%%%%%%%%%%%%%%%%%%%%%%%
\subsection{Sinh-Gordon model}\label{sec:ShGM}
%%%%%%%%%%%%%%%%%%%%%%%%%%%%%%%%%
The ShGM is one of the simplest non-trivial integrable quantum field theories. It contains only one real bosonic field $\phi(x)$, in terms of which the action reads~\cite{Mussardo10,Kormos-10}  
\begin{equation}
\label{LShG}
\mathcal{S} = \int \dd^2x\left[\frac{1}{2c^2}\left(\frac{\partial\phi}{\partial t}\right)^2-\frac{1}{2}\left(\frac{\partial\phi}{\partial x}\right)^2 + \frac{m_0^2c^2}{g^2}\bigl(1-\cosh g\phi\bigr)\right],
\end{equation}
where $m_0$ is the bare mass, $c$ the light velocity, and $g$ the dimensionless coupling constant.\footnote{We set $\hbar=1$ throughout the manuscript.} The model possesses an infinite number of conserved charges. Its spectrum is made up by spinless neutral particles with renormalised mass~\cite{BabujianKarowski02jpa} 
\begin{equation}
\label{ParticleMass}
m^2 = m_0^2\frac{\sin{\pi B}}{\pi B}, \quad B = \frac{cg^2}{8\pi+cg^2}.
\end{equation}
Introducing the rapidity $\theta$ to parametrise energy and momentum via $E=mc^2\cosh\theta$ and $p=mc\sinh\theta$, the two-particle scattering matrix is given by~\cite{Arinshtein-79}
\begin{equation}
\label{SShG}
S(\theta) = \frac{\sinh{\theta} - \ii\sin\pi B}{\sinh{\theta} + \ii\sin\pi B}.
\end{equation}
For $g=0$ and $g\to\infty$ we recover a free bosonic theory with the scattering matrix simply being $S(\theta)=+1$. The S-matrix is invariant under the duality transformation $g\to 8\pi/g$, which is not manifest at the level of the Lagrangian. The complete understanding of the duality still constitutes an open question~\cite{Konik-21,BernardLeClair22}. In this article we consider only $0\le g\le \sqrt{8\pi}$. Furthermore, the action is invariant under the $\mathbb{Z}_2$-transformation 
\begin{equation}
\label{Parity}
P: \phi(x)\to -\phi(x).
\end{equation}
The Hilbert space is built up by acting with the creation operators $Z^\dagger(\theta)$ on the vacuum state $\ket{0}$, thereby creating a particle with rapidity $\theta$. Together with the annihilation operators $Z(\theta)$ they satisfy the Faddeev--Zamolodchikov algebra, 
\begin{eqnarray}
\label{ZFA1}
Z(\theta)Z(\theta') &=& S(\theta-\theta')Z(\theta')Z(\theta),\\
\label{ZFA2}
Z^{\dagger}(\theta)Z^{\dagger}(\theta') &=& S(\theta-\theta')Z^{\dagger}(\theta')Z^{\dagger}(\theta),\\
\label{ZFA3}
Z(\theta)Z^{\dagger}(\theta') &=& S(\theta'-\theta)Z^{\dagger}(\theta')Z(\theta) + 2\pi\delta(\theta-\theta').
\end{eqnarray}
The algebra allows the exchange of particles in the asymptotic states. In the following, we order the rapidities both in in- and out-states in an increasing fashion. 

Once the S-matrix is known, form factors of operators can be obtained~\cite{Fring-93,KoubekMussardo93}. We are interested in two kinds of operators: the normal ordered field powers $:\phi^n:$ and vertex operators $e^{\alpha g\phi}$. Formally, the two are related as we consider the expansion of the second in powers of the parameter $\alpha$,
\begin{equation}
\label{FPvsVO}
e^{\alpha g\phi} = 1 + \sum_{n=1}^\infty\frac{\alpha^ng^n}{n!}A_n(\alpha,g):\phi^n:.
\end{equation}
Here the factors $A_n(\alpha,g)$ appear due to the necessary normal ordering. Given the parity invariance \eqref{Parity}, only the even terms are non-zero. The importance of this relation lies in the fact that form factors for vertex operator can be straightforwardly obtained, while in the field power case they are then found taking the $n$-th term of the power series. Moreover, the vertex operator for $\alpha=1$ is equivalent to the renormalisation-group relevant operator in the Lagrangian, i.e., $\cosh g\phi$.

An important tool in the study of integrable systems is provided by the TBA~\cite{Zamolodchikov90,Mussardo10}. It can be viewed as a quantisation condition on the rapidity spectrum of the excitations, since one has to ensure the consistency of the imposed boundary conditions with the two-particle scattering. For example, in a finite system of length $L$ with $N$ particles, the resulting Bethe equations read 
\begin{equation}
\label{BAShG}
mcL\sinh{\theta_j} = 2\pi I_j + \sum_{k=1, k\neq j}^N\chi{(\theta_k - \theta_j)}, \quad \chi(\theta) = -\ii\log{S(\theta)}.
\end{equation}
Here $I_j$ are positive integers since we are dealing with a bosonic theory. In the thermodynamic limit ($N, L\to\infty$, keeping $N/L$ constant) we can describe the occupied states (roots) and those that are not (holes) by their densities $\rho^{\mathrm{(r)}}(\theta)$ and $ \rho^{\mathrm{(h)}}(\theta)$, respectively. In terms of these the Bethe equations become 
\begin{equation}
\label{TBAdens}
\rho^{\mathrm{(r)}}(\theta) + \rho^{\mathrm{(h)}}(\theta) = \frac{mc}{2\pi}\cosh{\theta} + \int_{-\infty}^{\infty}\frac{\dd\theta'}{2\pi}\varphi(\theta-\theta')\rho^{\mathrm{(r)}}(\theta').
\end{equation}
The kernel $\varphi$ takes into account the scattering contributions and is explicitly given by 
\begin{equation}
\label{KernelShG}
\varphi(\theta) = \frac{\dd}{\dd\theta}\chi(\theta) = \frac{2\sin\pi B\,\cosh{\theta}}{\sinh^2{\theta}+\sin^2\pi B}.
\end{equation}
The solution is now obtained minimising the free energy, which yields the condition 
\begin{equation}
\label{TBAeq}
\log\eta(\theta) = mcR\cosh{\theta} - \int_{-\infty}^{\infty}\frac{\dd\theta'}{2\pi}\varphi(\theta-\theta')\log{\left(1+\frac{1}{\eta(\theta')}\right)},
\end{equation}
where $R=1/T$ with the temperature $T$, and we introduced 
\begin{equation}
\log\eta(\theta) = \frac{\rho^{\mathrm{(h)}}(\theta)}{\rho^{\mathrm{(r)}}(\theta)}.
\end{equation}
The coupled set of integral equations (\ref{TBAdens}) and (\ref{TBAeq}) form the TBA equations for the ShGM.

%%%%%%%%%%%%%%%%%%%%%%%%%%%%%%%%%
\subsection{Lieb--Liniger model}\label{Lieb-Liniger model}
%%%%%%%%%%%%%%%%%%%%%%%%%%%%%%%%%
The LLM is a non-relativistic interacting quantum field theory for the complex bosonic field $\Psi$. Using $\mu$ to denote the particle mass [as in (\ref{LShG})], the action of the system is given by~\cite{KorepinBogoliubovIzergin93}
\begin{equation}
\label{LLL}
\tilde \mathcal{S}= \int \dd^2x\left[\frac{\ii}{2}\left(\Psi\frac{\partial\Psi^\dagger}{\partial t} - \frac{\partial\Psi}{\partial t}\Psi^\dagger\right) -  \frac{1}{2\mu}\frac{\partial\Psi^\dagger}{\partial x}\frac{\partial\Psi}{\partial x} - \kappa\Psi^\dagger\Psi^\dagger\Psi\Psi\right].
\end{equation}
The coupling constant $\kappa$ tunes the interaction strength, depicted as a four-particle vertex, and it is considered to be positive. Here and in the following we use a tilde to distinguish objects related to the LLM from the corresponding quantities in the ShGM. 

The LLM can be described using the Bethe ansatz in a similar way to the ShGM discussed above. This allows one to find the spectrum of the theory~\cite{LiebLiniger63,Lieb63}, leading us to the non-relativistic factorisable S-matrix and the identification of an infinite set of conserved quantities, highlighting the integrability of the model. Specifically, the two-particle S-matrix is found to be
\begin{equation}
\label{SLL}
\tilde S(p) = \frac{p - 2\ii\mu\kappa}{p + 2\ii\mu\kappa},
\end{equation}
where $p$ is the momentum difference between the two particles. As for the previous case, the theory does not show any bound states as the interaction is repulsive.

The operators of interest in the quench setup are those that can be constructed from the field operators by taking powers like $\left(\Psi^\dagger\right)^n\Psi^n$. The reasons behind this are twofold: First, they can be considered as $n$-point correlation functions at $x=0$. Second, they encode the full counting statistics for the particle-number fluctuation in small intervals~\cite{Bastianello-18}.

Given the underlying Bethe ansatz structure, the thermodynamics of the model is available through the TBA equations. They are indeed built up in complete analogy to the relativistic case. Considering a system of length $L$ with $N$ particles and periodic boundary conditions, they read [cf. (\ref{BAShG})]
\begin{equation}
\label{BALL}
Lp_j = 2\pi \tilde I_j + \sum_{k=1, k\neq j}^N\tilde\chi{(p_k - p_j)},\quad \tilde\chi(p) = -\ii\log{\tilde S(p)},
\end{equation}
where as before $\tilde I_j$ are integers. Again introducing the density of roots $\tilde{\rho}^{\mathrm{(r)}}$ and holes $\tilde{\rho}^{\mathrm{(h)}}$ and their ratio $\tilde{\rho}^{\mathrm{(r)}}/\tilde{\rho}^{\mathrm{(h)}}=\log{\tilde\eta}$ and minimising the free energy (containing the Yang--Yang entropy~\cite{YangYang69} with $R=1/T$), one finds the set of equations
\begin{eqnarray}
\label{TBAdensLL}
\tilde\rho^{\mathrm{(r)}}(p) + \tilde\rho^{\mathrm{(h)}}(p) = \frac{1}{2\pi} + \int_{-\infty}^{\infty}\frac{\dd p'}{2\pi}\tilde\varphi(p-p')\tilde\rho^{\mathrm{(r)}}(p'),\\
\label{KernelShGLL}
\tilde\varphi(p) = \frac{\dd}{\dd p}\tilde\chi(p) = \frac{4\mu\kappa}{p^2+4\mu^2\kappa^2},\\
\label{TBAeqLL}
\log\tilde\eta(p) = R\frac{p^2}{2\mu} - \int_{-\infty}^{\infty}\frac{\dd p'}{2\pi}\tilde\varphi(p-p')\log{\left(1+\frac{1}{\tilde\eta(p')}\right)}.
\end{eqnarray}

As we have seen, the two models have many features in common. In the following subsection we discuss the  link between them in more detail~\cite{Kormos-10} .
As we will see, this will turn out to be extremely useful when computing form factors in the LLM.

%%%%%%%%%%%%%%%%%%%%%%%%%%%%%%%%%
\subsection{Non-relativistic limit}\label{Non-relativistic limit}
%%%%%%%%%%%%%%%%%%%%%%%%%%%%%%%%%
As was shown by Kormos et al.~\cite{Kormos-10}, performing a combination of non-relativistic and weak-coupling limits, the ShGM is mapped to the LLM. More precisely, taking the double limit 
\begin{equation}
\label{NRlimit}
c\to\infty ,\quad  g\to 0, \quad gc=4\sqrt{\kappa}=\mathrm{const.},
\end{equation}
the S-matrix of the ShGM (\ref{SShG}) is transformed into its LLM counterpart \eqref{SLL}. This requires the identification of the masses in both models, i.e., $m=\mu$, which in the limit \eqref{NRlimit} simplifies to $m_0=\mu$. Furthermore, we have to link the field operators of the two models, which is provided by the relation
\begin{equation}
\label{NRmapfield}
\phi(x,t) = \sqrt{\frac{1}{2\mu}}\left[e^{-\ii\mu c^2t}\Psi(x,t) + e^{\ii\mu c^2t}\Psi^{\dagger}(x,t)\right].
\end{equation}
This relation also allows the mapping of the two Lagrangians onto each other. In this work we are mostly concerned with the expectation values of field operators, 
\begin{equation}
\label{NRmapOEV}
\lim_{\mathrm{NR}}\langle :\phi^{2n}:\rangle ={2n\choose n}\left(\frac{1}{2\mu}\right)^n\langle\left(\Psi^\dagger\right)^n\Psi^n\rangle,
\end{equation}
where the subscript 'NR' denotes the non-relativistic limit. Using the expansion (\ref{FPvsVO}) one can show~\cite{BastianelloPiroli18} that the expectation values of vertex operators are formally mapped onto sums of expectation values of field powers, 
\begin{equation}
\label{NRmapVO}
\lim_{\mathrm{NR}}\langle e^{\alpha g\phi}\rangle = 1 + \sum_{n=1}^\infty\left[1-\cos(4a\kappa)\right]^n\frac{\langle\left(\Psi^\dagger\right)^n\Psi^n\rangle}{(\mu\kappa)^n n!^2},
\end{equation}
where we defined the parameter $a$ through the limiting procedure $\lim_{\mathrm{NR}} \alpha g = 4a\sqrt{\kappa}$. Using the same line of argument we can derive the form factors in the LLM from the ones in the ShGM, the latter known from the bootstrap approach~\cite{Kormos-10jstat}. The only condition we have to impose is that the expectation value of any state remains finite during the limiting process (regardless the operator we are considering), namely
\begin{eqnarray}
\label{NRmapFF}
&\lim_{\mathrm{NR}}c^{\frac{N+M}{2}}e^{\ii (N-M)\mu c^2t}F_{N+M}^{O}(\xi_1+\ii\pi,\dots,\xi_M+\ii\pi,\theta_1,\dots,\theta_N)\nonumber\\*
& \quad = \tilde F_{N+M}^{\tilde O}(p_1,\dots,p_M|q_1,\dots,q_N).
\end{eqnarray}
Here the operator $O$ is mapped onto its non-relativistic counterpart $\tilde O$, the rapidities $\xi_i,\theta_j$ turn into the non-relativistic momenta $p_i,q_j$, and their form factors are defined in the usual way, see \ref{Form Factors: axioms and properties} for more details. The non-relativistic nature of the LLM requires the strict distinction between the ingoing and outgoing momenta since crossing symmetry is absent.

%%%%%%%%%%%%%%%%%%%%%%%%%%%%%%%%%
\section{Quenching procedure and stationary state in the ShGM}
\label{Quenching procedure and stationary state}
%%%%%%%%%%%%%%%%%%%%%%%%%%%%%%%%%
This chapter we discuss the general setup under investigation, i.e., the considered quench protocol. In particular, we discuss the initial state for the time evolution and the eventual stationary (read final) state for the ShGM. We relate this to various results discussed in the literature~\cite{Sotiriadis-12,Sotiriadis-14,Bertini-16jsm}.

%%%%%%%%%%%%%%%%%%%%%%%%%%%%%%%%%
\subsection{Quench protocol}
\label{Quench protocol}
%%%%%%%%%%%%%%%%%%%%%%%%%%%%%%%%%
A natural way to realise a homogeneous quantum quench~\cite{CalabreseCardy06} is by changing abruptly some parameters of the considered model. In the ShGM we can use two parameters: the bare mass and the coupling constant. For instance, one could consider a quenching procedure that keeps the physical mass \eqref{ParticleMass} of the particles fixed, while changing the strength of their interactions. In such a quench, the energy spectrum would remain the same but the occupation numbers would be affected. Another class of interesting protocols are those that start or end from a free theory; there a fraction of conserved quantities are destroyed or restored by the quenching procedure itself.

However, the investigation of such a quench protocol is far from straightforward. In fact, it requires the determination of the initial state for the post-quench time evolution as function of the varying parameters. This has been achieved so far only in special cases, e.g., in Ising chains in transverse and longitudinal magnetic fields~\cite{Rossini-10,Hodsagi-18,Hodsagi-19}, the N\'{e}el state in the XXZ Heisenberg chain~\cite{Pozsgay14,Brockmann-14}, or the BEC state in the LLM~\cite{DeNardis-14,Ryland-22}. Although some attempts~\cite{Sotiriadis-14,Fagotti-14,Delfino14,S15,Horvath-16,DelfinoViti17,Piroli-17npb} have been made, a general framework to link the initial state to the quench protocol is still missing. We will not address this problem here, but instead consider a specific initial state in the following section. 

%%%%%%%%%%%%%%%%%%%%%%%%%%%%%%%%%
\subsection{Initial State}
\label{Initial State}
%%%%%%%%%%%%%%%%%%%%%%%%%%%%%%%%%
In this article we assume the initial state for the post-quench time evolution to be of the form of a squeezed coherent state 
\begin{equation}
\label{InitialSt}
|\psi\rangle = \exp{\left[\int_0^{\infty}\frac{\dd\theta}{2\pi}K(\theta)Z^\dagger(-\theta)Z^\dagger(\theta)\right]|}0\rangle.
\end{equation}
In this ansatz the pair amplitude $K$ contains all the information related to the quench protocol. For consistency it has to satisfy 
\begin{equation}
\label{KProp}
K(-\theta) = S(-2\theta)K(\theta).
\end{equation}
As was shown in References~\cite{Sotiriadis-14,Horvath-16}, for quenches from a system with large bare mass $m_0$ and vanishing interaction, the initial state indeed is very well represented by the squeezed coherent form with the pair amplitude being related to the boundary reflection amplitude $K_\mathrm{D}$ for Dirichlet boundary conditions. Motivated by these considerations, in this work  we use the initial state \eqref{InitialSt} with the pair amplitude 
\begin{equation}
\label{KSTM}
K(\theta) = K_\mathrm{D}(\theta)\,\frac{E_0(\theta)-E(\theta)}{E_0(\theta)+E(\theta)},
\end{equation}
where $E_0(\theta)=c^2\sqrt{m_0^2+m^2\sinh\theta}$ and $E(\theta)=m c^2\cosh\theta$, and the boundary reflection amplitude for Dirichlet boundary conditions reads~\cite{Ghoshal94}
\begin{equation}
\label{BoundaryK}
K_\mathrm{D}(\theta) = \ii\tanh\frac{\theta}{2}\,\frac{1+\cot{\frac{\pi B - 2\ii\theta}{4}}}{1-\tan{\frac{\pi B + 2\ii\theta}{4}}}.
\end{equation}
We note that at small rapidities the pair amplitude vanishes as $K(\theta)\sim\theta$. The initial state is translational invariant since it is built as the coherent sum of zero-momentum pairs, $K(\theta)|-\theta,\theta\rangle\equiv K(\theta)Z^\dagger(-\theta)Z^\dagger(\theta)\ket{0}$. They behave like non-interacting bosons under exchange [this can be directly proven from (\ref{SShG}) and (\ref{KProp})], even though the individual particles are interacting. Furthermore we note that the boundary reflection amplitude \eqref{BoundaryK} satisfies the boundary bootstrap relations~\cite{GhoshalZamolodchikov94}, which is also true for the pair amplitude \eqref{KSTM}.

In order to regularise the ultraviolet behaviour of the initial state, we further introduce another parameter $R$ via 
\begin{equation}
\label{ThermK}
K(R,\theta) = K(\theta)\,e^{-mcR\cosh{\theta}}.
\end{equation}
This acts as a cutoff in the production of particles at the characteristic momentum $p^*\sim 1/R$, while \eqref{KProp} remains valid. This simple way of introducing a cutoff was advocated in Reference~\cite{FiorettoMussardo10} in analogy to the extrapolation time $\tau_0\sim R/(2c)$ in critical systems~\cite{CalabreseCardy06,CalabreseCardy07}. In References~\cite{Sotiriadis-14,Horvath-16} it was shown, however, that a single extrapolation time is not sufficient to obtain universal results for different observables, which in fact requires the introduction of a momentum-dependent extrapolation time.

%%%%%%%%%%%%%%%%%%%%%%%%%%%%%%%%%
\subsection{Quench action method and gTBA equations}
\label{Quench Action method and gTBA equations}
%%%%%%%%%%%%%%%%%%%%%%%%%%%%%%%%%
The quench action method~\cite{CauxEssler13,Caux16} allows the calculation of the post-quench time evolution in integrable quantum models. We will review it in the next section; here we already present results for the steady state. We consider a system of length $L$ with $N$ particle pairs. The stationary state $|\rho\rangle$ must be an eigenstate of the Hamiltonian. In the thermodynamic limit the corresponding root density $\rho^{\mathrm{(r)}}(\theta)$ obeys a generalised version of the TBA (gTBA) equation, namely
\begin{equation}
\label{gTBA}
\log\eta(\theta) = -2\log{|K(\theta)|} - \int_{-\infty}^{\infty}\frac{\dd\theta'}{2\pi}\varphi(\theta-\theta')\log{\left(1+\frac{1}{\eta(\theta')}\right)},
\end{equation}
where $\eta(\theta)$ is related to the ratio of the hole and root densities via $\log\eta(\theta)=\rho^{\mathrm{(h)}}(\theta)/\rho^{\mathrm{(r)}}(\theta)$. Note that compared to the TBA equation (\ref{TBAeq}) a different driving term $-2\log{|K(\theta)|}$ appears. In addition, the condition \eqref{TBAdens} applies. Including the ultraviolet cutoff via \eqref{ThermK} the gTBA equations become
\begin{eqnarray}
\log\eta(\theta) &= &-2\log{|K(\theta)|} + 2mcR\cosh{\theta} \nonumber\\*
& & \qquad- \int_{-\infty}^{\infty}\frac{\dd\theta'}{2\pi}\varphi(\theta-\theta')\log{\left(1+\frac{1}{\eta(\theta')}\right)}.
\label{gTBAth}
\end{eqnarray}
An alternative way to obtain this result is given by the boundary TBA introduced in Reference~\cite{LeClair-95}. After a Wick rotation in the time direction, the system is put on a cylinder of radius $L$ and length $R$ with the boundary state placed at the imaginary times $y=0$ and $y=R$. By considering scattering along the $L$-channel, we use the resulting Bethe equations, which are modified by the presence of the boundaries, for the quantisation of the latter. Then taking the thermodynamic limit, $L\to\infty$, we arrive at (\ref{gTBAth}). The $R$ parameter can be related to the inverse of the temperature of the system when the quench is performed; a pure boundary function $K$ is related to infinite temperature solutions. 

This result is relevant for two reasons: First, it provides a natural cut-off for all integrals we are going to compute; namely we do not need to restrict our Hilbert space in order to obtain finite results and it simply relates to physical observables. Second, as it will turn out, it allows us to map our results to the non-relativistic regime where the number of particles is finite. Moreover this equivalence highlights once again the relation between the initial state of a quantum quench with certain boundary states.

%%%%%%%%%%%%%%%%%%%%%%%%%%%%%%%%%
\subsection{Thermal expectation values}
\label{Thermal expectation values}
%%%%%%%%%%%%%%%%%%%%%%%%%%%%%%%%%
In light of the previous considerations, it is natural to consider the stationary state as the thermal state of a boundary integrable field theory. Then the expectation value of an operator at infinite time is its thermal one, which can be computed via the LeClair--Mussardo formula~\cite{LeclairMussardo99} 
\begin{equation}
\label{LMseries}
\langle O\rangle_\mathrm{st} = \sum_{N=0}^{\infty}\int_{0}^{\infty}\prod_{j=1}^N\frac{\dd\theta_j}{2\pi}f(\theta_j)F_{2N,\mathrm{conn}}^O(-\theta_N,-\theta_{N-1},\dots,\theta_{N-1},\theta_N).
\end{equation}
The thermal state contributes through the filling fractions 
\begin{equation}
\label{FillFrac}
f(\theta) = \frac{1}{1+\eta(\theta)},
\end{equation}
while only the connected part of the form factors has to be considered. This can be extracted using the following limiting procedure 
\begin{eqnarray}
\label{FFconn}
&F_{2N,\mathrm{conn}}^O(\theta_1+\ii\pi,\dots,\theta_N+\ii\pi,\theta_1,\ldots,\theta_N)\nonumber\\
&\quad=\mathrm{FP}\left\{\lim_{\left\{\epsilon_i\right\}\to 0^+}F_{2N}^O(\theta_1 + \ii\pi + \ii\epsilon_1,\dots,\theta_N+ \ii\pi + \ii\epsilon_N,\theta_1,\dots,\theta_N)\right\},
\end{eqnarray}
where the script FP stands for the finite part of the limit, namely the part independent from any of the $\epsilon_i$. For later use we also state the filling fraction the non-relativistic regime,
	\begin{equation}
		\label{FillFracNR}
		\tilde f(p) = \frac{1}{\tilde{\eta}(p)+1}.
\end{equation}

The procedure of computing the expectation values from the LeClair--Mussardo series can be simplified when considering vertex operators. An equivalent expression, derived by Negro and Smirnov~\cite{NegroSmirnov13}, has been proven to be useful also for the quench case~\cite{Bertini-16jsm}. Given (\ref{FPvsVO}) we are able to retrieve the expectation values of the field operators $:\phi^{2n}:$ from the vertex operator one; the Negro--Smirnov formula then reads 
\begin{eqnarray}
	\label{NSform}
	&\frac{\langle e^{(\alpha+1)g\phi}\rangle_\mathrm{st}}{\langle e^{\alpha g\phi}\rangle_\mathrm{st}} = \frac{\mathcal{G}(\alpha+1)}{\mathcal{G}(\alpha)}\!\left[1 + \frac{2\sin{(\pi B(2\alpha+1))}}{\pi}\int_{-\infty}^{\infty}\frac{\dd\theta\,e^\theta}{1+\eta(\theta)}p_\alpha(\theta)\right]\!, \\
	&p_\alpha(\theta) = e^{-\theta} + \int_{-\infty}^{\infty}\frac{\dd\xi}{1+\eta(\xi)}\chi_\alpha(\theta-\xi)p_\alpha(\xi),\\
&\chi_\alpha(\theta) = \frac{\ii}{2\pi}\left(\frac{e^{-2\pi\ii\alpha B}}{\sinh{(\theta+\ii\pi B)}} - \frac{e^{2\pi\ii\alpha B}}{\sinh{(\theta-\ii\pi B)}}\right).
\end{eqnarray}
Hence, as described in Reference~\cite{Bertini-16jsm}, we can recover the expectation value of the vertex operator for any value of $\alpha$ and, from that, for any field power operator.

%%%%%%%%%%%%%%%%%%%%%%%%%%%%%%%%%
\section{Post-quench time evolution}
\label{Post-quench time evolution chap}
%%%%%%%%%%%%%%%%%%%%%%%%%%%%%%%%%
The goal of this section is the computation of the time evolution of the observables in the ShGM. We calculate them using two different formalisms, the previously introduced quench action method and the linked cluster expansion; we find agreement between the two. Then the late time dynamics is obtained and discussed. The derivation in the main text will be for the normal ordered squared field operator $:\phi^2:$, but results will also be stated for $:\phi^4:$ and the vertex operator. Technical details of the computation are presented in \ref{Linked Cluster expansion for Sinh-Gordon model} and \ref{Quench Action method for Sinh-Gordon model}. 

%%%%%%%%%%%%%%%%%%%%%%%%%%%%%%%%%
\subsection{Linked cluster expansion}
\label{Linked Cluster expansion}
%%%%%%%%%%%%%%%%%%%%%%%%%%%%%%%%%
An intuitive way to describe the post-quench dynamics consists in considering the so-called small quench regime, i.e., those initial states which are close (in the sense of expectation values) to the post-quench ground state. We can see that this is equivalent to requiring that the pair amplitude is small, $K\ll 1$, such that we can expand the exponential over the number of particles. This is the linked cluster expansion~\cite{FiorettoMussardo10,KormosPozsgay10,Calabrese-11,SE12,Castro-Alvaredo-19,Castro-Alvaredo-20}. We can also justify this by multiplication of the pair amplitude with a constant $\epsilon$, which does not affect the relevant constraints on $K(\theta)$ and thus can be used as a formal expansion parameter. However, following the existing literature, we shall consider the pair amplitude directly as the expansion parameter. 

Our goal is the derivation of a formal expansion for the one-point function
\begin{equation}
\label{LC}
\lim_{L\to\infty}\left[\frac{\langle\psi| O(t) |\psi\rangle}{\langle\psi|\psi\rangle}\right]_L = \lim_{L\to\infty}\left[\frac{\langle\psi| e^{\ii Ht} O e^{-\ii Ht}|\psi\rangle}{\langle\psi|\psi\rangle}\right]_L,
\end{equation}
where $H$ denotes the Hamiltonian governing the post-quench dynamics, and the subscript $L$ stands for the finite-volume regime. The expansion of the numerator gives 
\begin{eqnarray}
\label{NumExp}
\langle\psi| e^{\ii Ht} O e^{-\ii Ht}|\psi\rangle_L&=& \sum_{M,N=0}^{\infty}\frac{1}{M!N!}\int_{0}^{\infty}\prod_{a=1}^{M}\frac{\dd\xi_a}{2\pi}K^*(\xi_a)\prod_{b=1}^{N}\frac{\dd\theta_b}{2\pi}K(\theta_b)\nonumber\\*
& &\times \langle -\xi_M,-\xi_{M-1},\dots,\xi_M|O|-\theta_N,\dots,\theta_N\rangle_L\nonumber\\*
& &\times e^{2\ii mc^2t\left(\sum_{a=1}^M\cosh{\xi_a}-\sum_{b=1}^N\cosh{\theta_b}\right)}\nonumber\\
& =& \sum_{M,N=0}^{\infty} C_{M,N}^O(t),
\end{eqnarray}
while the denominator becomes
\begin{eqnarray}
\label{DenExp}
\langle\psi|\psi\rangle_L &=&\sum_{N=0}^{\infty}\frac{1}{N!^2}\int_{0}^{\infty}\prod_{a=1}^{N}\frac{\dd\xi_a}{2\pi}\frac{\dd\theta_a}{2\pi}K^*(\xi_a)K(\theta_a)\nonumber\\*
& &\times \langle -\xi_N,-\xi_{N-1},\dots,\xi_N|-\theta_N,\dots,\theta_N\rangle_L\nonumber\\
&= &\sum_{N=0}^{\infty} Z_{2N}.
\end{eqnarray}
Thus we obtain the formal expansion
\begin{equation}
\label{InvDenExp}
\frac{1}{\langle\psi|\psi\rangle_L} = 1-Z_2+Z_2^2-Z_4+\mathcal{O}(K^6),
\end{equation}
which identifies the diverging parts in the infinite-volume limit, namely the linked clusters. Both the numerator and the denominator contain divergencies, but the expectation values do not; hence all divergencies have to cancel when the product between (\ref{NumExp}) and (\ref{InvDenExp}) is taken. The finite-size regularisation scheme allows us to write down all the divergent terms as proportional to a positive power of the system size $L$, meaning that, prior to the limiting procedure, the resulting terms $D_{M,N}(t)$ are finite. The divergencies-free series will then be written as
\begin{equation}
\label{LC2}
\lim_{L\to\infty}\left[\frac{\langle\psi| O(t) |\psi\rangle}{\langle\psi|\psi\rangle}\right]_L = \frac{\sum_{M,N=0}^{\infty} C_{M,N}^O(t)}{\sum_{N=0}^{\infty} Z_{2N}} = \sum_{M,N=0}^{\infty} D_{M,N}^O(t).
\end{equation}
We are going to compute this expression for the considered operators up to the next-to-leading order.

%%%%%%%%%%%%%%%%%%%%%%%%%%%%%%%%%
\subsection{Quench action method}
\label{Quench Action method}
%%%%%%%%%%%%%%%%%%%%%%%%%%%%%%%%%
As previously mentioned, the quench action method is able to capture the properties of the stationary state $|\rho\rangle$ (or representative state) reached via the post-quench dynamics. Its particle distribution is found by the means of gTBA equations (\ref{gTBAth}). Furthermore, information about the time evolution of the system can be extracted from the representative state itself: considering an integrable theory in a finite volume $L$, it was argued that~\cite{CauxEssler13,Caux16}
\begin{eqnarray}
\lim_{L\to\infty}\left[\frac{\langle\psi| O(t) |\psi\rangle}{\langle\psi|\psi\rangle}\right]_L &=& \frac{1}{2}\lim_{L\to\infty}\left[\frac{\langle\psi|O(t) |\rho\rangle}{\langle\psi|\rho\rangle} + \frac{\langle\rho|O(t) |\psi\rangle}{\langle\rho|\psi\rangle}\right]_L \nonumber\\
&=&\lim_{L\to\infty}\mathfrak{Re}\left[\frac{\langle\psi|O(t) |\rho\rangle}{\langle\psi|\rho\rangle}\right]_L.
\label{QAM}
\end{eqnarray}
The normalisation adopted for the representative state is given in terms of its particle density
\begin{equation}
\label{RhoNorm}
\langle\rho|\rho\rangle = \lim_{L\to\infty} L\int_{0}^{\infty}\frac{\dd\theta}{2\pi}\rho^\mathrm{(r)}(\theta).
\end{equation}
Further details regarding integrable theories in finite volume are revisited in \ref{Sinh-Gordon model in finite volume}. Eventually, it was shown that, taking the limit $t\to\infty$ and applying the saddle point approximation, the expression (\ref{QAM}) becomes
\begin{equation}
\label{StOEV}
\lim_{t\to\infty}\lim_{L\to\infty}\left[\frac{\langle\psi| O(t) |\psi\rangle}{\langle\psi|\psi\rangle}\right]_L = \lim_{L\to\infty}\left[\frac{\langle\rho| O |\rho\rangle}{\langle\rho|\rho\rangle}\right]_L,
\end{equation}
recovering the results discussed above.

The general strategy to compute the right-hand-side of (\ref{QAM}) can be divided into different steps:
\begin{enumerate}
\item Compute the denominator, i.e., the overlap $\langle\rho|\psi\rangle$. It is of order $\mathcal{O}(K^N)$, where $N$ is the number of particle pairs in the finite-volume representation.
\item Write down explicitly the numerator: the time dependence is solely given by exponentials. The analytic structure of the form factors contains, in the connected part, double poles from the annihilation pole axiom; when one of them is picked, the derivative over the rapidity gives a time contribution. We shall then neglect other contributions from the residue of the double pole and only keep the time dependent terms.
\item Then, in the sum over the number of particle pairs contained in the initial state, we can find the lowest order in $K$ in the expansion of the initial state that allows to extract the $N$ double poles contained in the form factors. For instance, considering the operator $:\phi^2:$, the annihilation pole axiom tells us that this state is the one labelled by $M=N+1$.
\item At this stage we can explicitly pick the poles, provided the expressions are correctly regularised in the finite volume, and afterward perform the ratio between numerator and denominator.
\item Finally, the thermodynamic limit can be taken.
\end{enumerate}

Our results will rely on the solution of (\ref{gTBAth}). This solution cannot be derived analytically, except in the small-quench regime. In this limit, we are able to obtain an explicit result for $\eta$, since the driving term only contributes to the lowest order of the expansion,
\begin{equation}
\label{EtaSol}
\eta^{-1}(\theta) = |K(\theta)|^2e^{-mcR\cosh{\theta}},
\end{equation}
while the particle density is obtained from (\ref{TBAdens}), again truncated from the integral part which gives higher-order contributions. We write it using the filling fraction defined in (\ref{FillFrac}) in order to make the relation to the particle density in the small-quench regime more transparent,
\begin{equation}
\label{PartDensSol}
\rho^\mathrm{(r)}(\theta) = \frac{mc}{2\pi}f(\theta)\cosh{\theta}=\frac{mc}{2\pi}\frac{|K(\theta)|^2}{e^{mcR\cosh{\theta}}+|K(\theta)|^2}\cosh{\theta}.
\end{equation}
This expression is very similar to the one one finds for free theories~\cite{Mussardo-17}, where the pair amplitude $K$ function plays the role of a generalised fugacity.

%%%%%%%%%%%%%%%%%%%%%%%%%%%%%%%%%
\subsection{Results for the late-time dynamics}
\label{Results and late-time dynamics}
%%%%%%%%%%%%%%%%%%%%%%%%%%%%%%%%%
Both approaches, the linked cluster expansion as well as the quench action method, yield a lot of terms that are not of particular interest. Specifically, they describe fast oscillating excitations or early-time dynamics, while we focus on the relaxation of the system at late times. Hence below we will only present the interesting terms in this sense. Using the linked cluster expansion (see \ref{Linked Cluster expansion for Sinh-Gordon model}) we obtain for the time evolution of the expectation value of the squared field operator 
\begin{eqnarray}
\label{LCphi2}
\langle :\phi^2:\rangle (t) &=& 2(1-\Gamma t)\,\mathfrak{Re}\int_0^{\infty}\frac{\dd\theta}{2\pi}K(\theta)F_2^{:\phi^2:}(-\theta,\theta)e^{-2\ii mc^2t\cosh{\theta}} \nonumber\\
& &+ \int_0^{\infty}\frac{\dd\theta}{2\pi}|K(\theta)|^2 F_{4,\mathrm{conn}}^{:\phi^2:}(-\theta+\ii\pi,\theta+\ii\pi,-\theta,\theta) + \ldots, 
\end{eqnarray}
where the connected form factor was defined in \eqref{FFconn}, and the dots represent contributions of $\mathcal{O}(K^3)$ as well as uninteresting terms (in the sense mentioned above) in lower orders. Similarly, for the fourth power of the field operators we find
\begin{eqnarray}
\label{LCphi4}
\langle :\phi^4:\rangle (t) &=&(1-\Gamma t)\int_0^{\infty}\frac{\dd\xi}{2\pi}\frac{\dd\theta}{2\pi}K^*(\xi)K(\theta) \nonumber\\*
&&\qquad\qquad\times F_4^{:\phi^4:}(-\xi+\ii\pi,\xi+\ii\pi,-\theta,\theta)\,e^{2\ii mc^2t(\cosh{\xi}-\cosh{\theta})} \nonumber\\
&& + 2(1-\Gamma t)\,\mathfrak{Re}\int_0^{\infty}\frac{\dd\xi_1}{2\pi}\frac{\dd\xi_2}{2\pi}K^*(\xi_1)K^*(\xi_2)\nonumber\\*
&& \qquad\qquad\times F_4^{:\phi^4:}(-\xi_2,-\xi_1,\xi_1,\xi_2)\,e^{2\ii mc^2t(\cosh{\xi_1}+\cosh{\xi_2})} \nonumber\\
&&+ \int_0^{\infty}\frac{\dd\theta}{2\pi}|K(\theta)|^2F_\mathrm{4,conn}^{:\phi^4:}(-\theta+\ii\pi,\theta+\ii\pi,-\theta,\theta) + \dots
\end{eqnarray}
where here the dots represent contributions of $\mathcal{O}(K^5)$ as well as uninteresting terms. We notice that the fourth power of the field operator shows a richer spectrum of oscillations than the quadratic one. This is due to the fact that at leading order the operator $:\phi^4:$ creates more excitations than $:\phi^2:$. Finally, for the vertex operator we obtain
\begin{eqnarray}
\label{LCexp}
\langle e^{\alpha g\phi}\rangle (t) &=& \mathcal{G}(\alpha) +  \int_0^{\infty}\frac{\dd\xi}{2\pi}|K(\xi)|^2F_{4,\mathrm{conn}}^\alpha(-\xi+\ii\pi,\xi+\ii\pi,-\xi,\xi)\nonumber\\
& &+ 2\,\mathfrak{Re}\int_0^{\infty}\frac{\dd\theta}{2\pi}K(\theta)F_2^\alpha(-\theta,\theta)e^{-2\ii mc^2t\cosh{\theta}} \nonumber\\
& &+ 2\,\mathfrak{Re}\int_0^{\infty}\frac{\dd\theta_1}{2\pi}\frac{\dd\theta_2}{2\pi}K(\theta_1)K(\theta_2)\nonumber\\*
&& \qquad\times F_4^\alpha(-\theta_2,-\theta_1,\theta_1,\theta_2)e^{-2\ii mc^2t(\cosh{\theta_1}+\cosh{\theta_2})}+\ldots,
\end{eqnarray}
with $\mathcal{G}(\alpha)=\bra{0}e^{\alpha\phi}\ket{0}$, and the dots representing contributions of $\mathcal{O}(K^3)$ as well as uninteresting terms. We note especially the appearance of an operator-independent relaxation rate (see below), namely 
\begin{equation}
\label{Gamma}
\Gamma = \frac{2mc^2}{\pi}\int_0^{\infty}\dd\theta|K(\theta)|^2\sinh{\theta}+\mathcal{O}(K^3).
\end{equation}
Moreover we are also able to describe the stationary one-point functions through the linked cluster formalism, since also time-independent values are obtained by means of connected form factors.

Now we turn to the calculation of the time evolution using the quench action method. The details of the derivation are given in \ref{Quench Action method for Sinh-Gordon model}, the final results read
\begin{eqnarray}
\langle:\phi^2: \rangle (t) &=& 2e^{-\Gamma t}\,\mathfrak{Re}\int_0^{\infty}\frac{\dd\xi}{2\pi} K(\xi)F_2^{:\phi^2:}(-\xi,\xi)e^{-2\ii mc^2t\cosh{\xi}}\nonumber\\
&&+ \int_0^{\infty}\frac{\dd\theta}{2\pi}|K(\theta)|^2 F_{4,\mathrm{conn}}^{:\phi^2:}(-\theta+\ii\pi,\theta+\ii\pi,-\theta,\theta),\label{QAMphi2}\\
\langle :\phi^4:\rangle (t)&=&e^{-\Gamma t}
\int_0^{\infty}\frac{\dd\xi}{2\pi}\frac{\dd\theta}{2\pi}K^*(\xi)K(\theta)\nonumber \\*
&&\qquad\times F_4^{:\phi^4:}(-\xi+\ii\pi,\xi+\ii\pi,-\theta,\theta)e^{2\ii mc^2t(\cosh{\xi}-\cosh{\theta})}\nonumber\\
&&+ 2e^{-\Gamma t}\,\mathfrak{Re}\int_0^{\infty}\frac{\dd\xi_1}{2\pi}\frac{\dd\xi_2}{2\pi}K^*(\xi_1)K^*(\xi_2)\nonumber\\*
&& \qquad\times F_4^{:\phi^4:}(-\xi_2,-\xi_1,\xi_1,\xi_2)\,e^{2\ii mc^2t(\cosh{\xi_1}+\cosh{\xi_2})}\nonumber\\
&&+ \int_0^{\infty}\frac{\dd\theta}{2\pi}|K(\theta)|^2F_\mathrm{4,conn}^{:\phi^4:}(-\theta+\ii\pi,\theta+\ii\pi,-\theta,\theta),  \label{QAMphi4}\\
\langle e^{\alpha g\phi}\rangle (t) &=&\mathcal{G}(\alpha) +    2e^{-\Gamma t}\,
\mathfrak{Re}\int_0^{\infty}\frac{\dd\theta}{2\pi}K(\theta)F_2^\alpha(-\theta,\theta)e^{-2\ii mc^2t\cosh{\theta}}\nonumber\\
&& + \int_0^{\infty}\frac{\dd\xi}{2\pi}|K(\xi)|^2F_\mathrm{4,conn}^\alpha(-\xi+\ii\pi,\xi+\ii\pi,-\xi,\xi).\label{QAMexp}
\end{eqnarray}
First, we note that expanding this result in $K$ returns the corresponding terms in the linked cluster expansion \eqref{LCphi2}--\eqref{LCexp}. Second, $\Gamma$ turns out to be the relaxation rate of the system, being operator independent and characterising the exponential suppression in the late-time dynamics. Hence the out-of-equilibrium fluctuations decay with a relaxation rate proportional to the mass of the excitations; a behaviour that has been observed in other massive field theories as well~\cite{Calabrese-11,SE12,BSE14,CS17,Castro-Alvaredo-19}. Third, we explicitly see that the formal expansion for the vertex operator (\ref{FPvsVO}) is applicable in the quench regime. This was of course expected since in integrable theories we are able to describe out of equilibrium dynamics using equilibrium field theory tools. Finally, the imaginary exponentials are responsible for the oscillations of the observables in the time evolution; here, the frequency of the oscillations is proportional to the mass of the lowest relevant excitation.

From the results (\ref{QAMphi2})--(\ref{QAMexp}) we can also extract the late-time dynamics by a saddle point approximation, yielding exponential decay with an operator-dependent power-law correction depending on the behaviour of the pair amplitude \eqref{KSTM} and form factors (see \ref{Form Factors: axioms and properties}) at small rapidities. Explicitly we obtain
\begin{eqnarray}
\label{LateTime}
\langle :\phi^2:\rangle (t)&\sim &2t^{-3/2}e^{-\Gamma t}\cos{(2mc^2t)} + \langle :\phi^2:\rangle_\mathrm{st},\\
\langle :\phi^4:\rangle (t)&\sim &t^{-3}e^{-\Gamma t} \left[1 + 2\cos{(4mc^2t)}\right]+ \langle :\phi^4:\rangle_\mathrm{st},\label{eq:phi4atationaryphase}\\
\langle e^{\alpha g\phi}\rangle (t)&\sim &2t^{-3/2}e^{-\Gamma t}\cos{(2mc^2t)} + \langle e^{\alpha g\phi} \rangle_\mathrm{st}.
\end{eqnarray}
This also reflects the different contributions of the diagonal and off-diagonal terms in the linked cluster approach: the former yield monotonically decaying terms, while the latter give rise to oscillations. In the next section we are going to relate these results to the dynamics in the LLM.

%%%%%%%%%%%%%%%%%%%%%%%%%%%%%%%%%
\section{Non-relativistic limit of quenches in the ShGM to ones in the LLM}
\label{Lieb-Liniger quenches as non-relativistic Sinh-Gordon ones}
%%%%%%%%%%%%%%%%%%%%%%%%%%%%%%%%%
In this section we take the non-relativistic limit of the results presented above, thus describing quenches in the LLM. In particular, we extract the counting statistics in the stationary state and the late-time dynamics of observables.

%%%%%%%%%%%%%%%%%%%%%%%%%%%%%%%%%
\subsection{Out-of-equilibrium non-relativistic limit}
\label{Out of equilibrium non-relativistic limit}
%%%%%%%%%%%%%%%%%%%%%%%%%%%%%%%%%
The non-relativistic limit from the ShGM to the LLM is well-established at equilibrium~\cite{Kormos-10}, e.g., at the level of the form factors, see (\ref{NRmapFF}). Since these constitute the essential input for our description of the non-equilibrium dynamics after a quantum quench, the non-relativistic limit should also be applicable for such situations. Thus we can map also the states, in the sense that this preserves the relation between relativistic and non-relativistic $n$-point functions.

The non-relativistic limit of the squeezed coherent state \eqref{InitialSt} becomes
\begin{equation}
\label{NRmapIS}
|\tilde\psi\rangle = \lim_{\mathrm{NR}}\ket{\psi}
=\exp{\left[\int_0^{\infty} \frac{\dd p}{2\pi} \tilde K(R,p) \tilde Z^{\dagger}(-p) \tilde Z^{\dagger}(p)\right]}|0\rangle.
\end{equation}
We note that this state is a coherent superposition of states with different particle numbers, and as such appears artificial in the usual setup of the LLM where the particle number is conserved. However, in order to consider comparable quench setups in the ShGM and LLM, and thus facilitate our study of the non-relativistic limit, we have to restrict us to initial states that are linked by the limit as well. Using the squeezed coherent state \eqref{InitialSt} in the ShGM, which allows the analysis of the quench dynamics as discussed above, then forces the consideration of the state \eqref{NRmapIS}. Here the integration is over the momenta $p$, the operators $\tilde Z(p)$ denote the Faddeev--Zamolodchikov operators acting in the Hilbert space of the LLM, and $\tilde K$ is straightforwardly obtained by the non-relativistic limit of the pair amplitude including the regularisation parameter $R$. In our case, (\ref{ThermK}) becomes
\begin{equation}
	\label{NRK}
	\tilde{K}(R,p) = \frac{\mu_0-\mu}{\mu_0+\mu}\frac{p}{p+\ii\mu\kappa}e^{-R\frac{p^2}{2\mu}}. 
\end{equation}
We note that the first factor on the right-hand side implies that non-trivial dynamics in the LLM is only possible if the bare mass in the ShGM is quenched. It can be straightforwardly shown that the boundary relations (\ref{BoundaryK}) are still valid after the limit under the mapping of the S-matrix itself. We stress that the quenching procedures have to be consistently related via the non-relativistic limit, i.e., the ShGM quench $m_0, g_0\to m,g$ is related to its LLM counterpart $\kappa_0\to\kappa$ under the constraint
\begin{equation}
\lim_{\mathrm{NR}}\frac{g_0}{g}=\sqrt{\frac{\kappa_0}{\kappa}}.
\end{equation}
We recall that the masses in both models are identified under the limiting procedure.

%%%%%%%%%%%%%%%%%%%%%%%%%%%%%%%%%
\subsection{Representative state and counting statistics} 
\label{Representative state and counting statistics}
%%%%%%%%%%%%%%%%%%%%%%%%%%%%%%%%%
We are now in a position to study the stationary state, starting from the non-relativistic gTBA equations. We note that the existence of the cutoff $R$ ensures a proper regularisation of the high-energy contributions. In contrast, without $R$ the solutions would be ill-defined, since the particle density at high momenta would tend to a finite value~\cite{Bertini-16jsm} corresponding to infinite temperatures.

The gTBA equation [together with (\ref{TBAdensLL}) and (\ref{KernelShGLL})] for the LLM reads 
\begin{eqnarray}
\log\tilde\eta(p) &=&-\log\frac{\mu_0-\mu}{\mu_0+\mu} -2\log{\frac{p^2}{p^2+\mu^2\kappa^2}}\nonumber\\
& &+\frac{Rp^2}{2\mu} - \int_{-\infty}^{\infty}\frac{\dd p'}{2\pi}\tilde\varphi(p-p')\log{\left(1+\frac{1}{\tilde\eta(p')}\right)}.
\label{gTBAthLL}
\end{eqnarray}
In this expression, the role of the first term on the right-hand side can be deduced straightforwardly: it determines the effect of the chemical potential in post-quench dynamics, as introduced in~\cite{Bertini-16jsm}; in other words, quenching the mass in the non-relativistic case is equivalent to introducing a non-zero chemical potential. Another effect of the chemical potential in the non-relativistic system is the cancellation of the rest mass of the particles~\cite{Kormos-10}. In the small quench regime we can compare this expression with the one for the ShGM 
\begin{equation}
\label{EtaSolLL}
\tilde\eta^{-1}(p) = |\tilde K(p)|^2e^{-R\frac{p^2}{2\mu}}.
\end{equation}
The consistency for the other thermodynamic quantities directly follows from this. For sake of completeness we report here the particle density given by 
\begin{equation}
\label{PartDensSolLL}
\tilde{\rho}^{\mathrm{(r)}}(\theta)=\frac{1}{2\pi}\frac{|\tilde K(p)|^2}{e^{R\frac{p^2}{2\mu}}+|\tilde K(p)|^2}.
\end{equation}

In the stationary state we are able to compute the expectation values of the operators $\Psi^n(\Psi^\dagger)^n$ for any $n$. This has been done~\cite{BastianelloPiroli18} by the means of the Negro--Smirnov formula (\ref{NSform}) and we strictly follow the same procedure; in the out-of-equilibrium framework we must use the particle density distribution  (\ref{PartDensSolLL}) given by the thermal gTBA equations, instead of the usual thermal ones. We obtain
\begin{equation}
\label{EVn}
\langle \Psi^n(\Psi^\dagger)^n\rangle_{\mathrm{st}} = (n!)^2(\mu\kappa)^n\sum_{\{n_j\}}\prod_j\left[\frac{1}{n_j!}\left(\frac{B_j}{2\pi \mu\kappa}\right)^{n_j}\right],
\end{equation}
where the sum is carried over all sets of strictly positive integers $n_j$ satisfying the constraint $\sum_jjn_j=n$, and the coefficients $B_j$ are related to the filling fractions (\ref{FillFracNR}) via
\begin{equation}
\label{Bj}
B_j = \frac{1}{j}\int_{0}^{\infty}\dd p\tilde f(p)b_{2j-1}(p).
\end{equation}
Here the auxiliary functions $b_j$ are determined by the following set of integral equations,
\begin{eqnarray}
\label{bj}
b_{2l}(p) &=& \int_{0}^{\infty}\frac{\dd q}{2\pi}\tilde f(q)\{\tilde\varphi(p-q)[b_{2l}(q)-b_{2l-2}(q)]\nonumber\\*
& &\qquad\qquad\qquad+ \Lambda(p-q)[b_{2l-1}(q)-b_{2l-3}(q)]\},\\
b_{2l+1}(p) &=& \delta_{l,0} + \int_{0}^{\infty}\frac{\dd q}{2\pi}\tilde f(q)\{\tilde\varphi(p-q)[b_{2l+1}(q)-b_{2l-1}(q)]\nonumber\\*
& &\qquad\qquad\qquad+ \Lambda(p-q)b_{2l}(q)\},\\
\Lambda(p) &=& \frac{2p}{p^2 + (2\mu\kappa)^2}.
\end{eqnarray}
With this we are in principle able to compute the expectation values of any function of the field operator in the stationary state. For example, an interesting quantity related to them is the full counting statistics of the number of particles within a small spatial interval of length $\Delta$: we introduce the operator $\hat{N}_\Delta$ that counts the number of particles in the interval, 
\begin{equation}
	\hat N_{\Delta} = \int_{0}^{\Delta}\dd x\Psi^{\dagger}(x)\Psi(x).
\end{equation}
This can be decomposed making use of the projection operator $\hat P_n$ on the subspace containing $n$ particles as
\begin{equation}
	\hat N_{\Delta} = \sum_{n=0}^{\infty}n\hat P_n.
\end{equation}
Now the expectation value $P_\Delta(n)=\bra{\Omega}\hat P_n\ket{\Omega}$, with $\ket{\Omega}$ a macroscopic state, can be linked~\cite{BastianelloPiroli18} to the field operator expectation value as 
\begin{equation}
\label{FCS}
P_\Delta(n) = \frac{\Delta^n}{n!}\left(\langle(\Psi^\dagger)^n\Psi^n\rangle_{\mathrm{st}} + \mathcal{O}(\Delta)\right).
\end{equation}
Hence we can write Taylor expansion of $P_n(\Delta)$ in terms of the expectation values.

%%%%%%%%%%%%%%%%%%%%%%%%%%%%%%%%%
\subsection{Post-quench time evolution}
\label{Post-quench time evolution sub}
%%%%%%%%%%%%%%%%%%%%%%%%%%%%%%%%%
As previously mentioned, the time evolution of the LLM can be obtained in three different ways: (i) by taking the non-relativistic limit of the results obtained for the ShGM, (ii) via the linked cluster expansion directly in the LLM starting from (\ref{NRmapIS}) or (iii) applying the quench action method. Here we focus on the non-relativistic limit and check the obtained results with linked cluster expansion computations. The latter approach is presented in \ref{Linked Cluster expansion for Lieb-Liniger model}.

The first quantity to derive is the relaxation rate $\tilde\Gamma$. This is directly obtained from (\ref{Gamma}) with the result 
\begin{equation}
\label{NRmapGamma}
\tilde\Gamma = \lim_{\mathrm{NR}}\Gamma = \frac{2}{\mu\pi}\int_0^{\infty}\dd p|\tilde K(p)|^2p.
\end{equation}
We note that the appearance of a finite relaxation rate is a consequence of the presence of annihilation poles in the non-relativistic form factors (as explicitly derived in Reference~\cite{IzerginKorepin84}). Another decisive property of the form factors of the field operator powers $(\Psi^\dagger)^n\Psi^n$ is the absence of off-diagonal terms, i.e., the number of incoming particles has to equal the number of outgoing ones, which leads to a drastic simplification of many expressions. The one-point function $\langle\Psi^\dagger\Psi\rangle$ is obtained directly from the squared field operator $\langle:\phi^2:\rangle$, see (\ref{LCphi2}) and (\ref{QAMphi2}). All off-diagonal terms vanish after taking the non-relativistic limit, as do the  diagonal time dependent ones given that form factors inside of the integrals are vanishing (see \ref{Linked Cluster expansion for Lieb-Liniger model}). Thus the remainder is given by the stationary contributions (the connected form factors do not vanish, as proven in~\cite{Kormos-10}) and we find $\langle\Psi^\dagger\Psi\rangle (t) = \mathrm{const.}\equiv n_{\mathrm{density}}$. This can be interpreted as that the mean particle density in the system remains constant during the relaxation process. The expectation values $\langle(\Psi^\dagger)^n\Psi^n\rangle$, $n>1$, as well as their generating function, i.e., the non-relativistic limit of the vertex operator, have a non-trivial time evolution. For example, for $n=2$ we get (omitting the stationary terms)
\begin{eqnarray}
\label{LLpsipsi2}
\langle(\Psi^\dagger)^2\Psi^2\rangle &=& e^{-\tilde\Gamma t}\int_0^{\infty}\frac{\dd p}{2\pi}\frac{\dd q}{2\pi}\tilde K^*(p)\tilde K(q)\nonumber\\*
& &\qquad\qquad\times\tilde F^{(\Psi^\dagger)^2(\Psi)^2}_4(-p,p|-q,q)e^{\ii\frac{t}{\mu}(p^2-q^2)}.
\end{eqnarray}
In this case the form factor inside the integral does not vanish and contains dynamical poles.

In addition to the relaxation rate, we can also extract the multiplicative power-law correction via a stationary phase approximation. Keeping in mind that both the form factors and the pair amplitude vanish linearly at small values of the momenta, we obtain the power law $t^{-3(N+M)/2}$ with $N$ and $M$ denoting the number of ingoing and outgoing particles (except for $\Psi^\dagger\Psi$). Since only diagonal terms contribute, $N=M$, we find in particular 
\begin{equation}
\langle\Psi^\dagger\Psi\rangle (t) \sim \mbox{const}, \quad \langle(\Psi^\dagger)^2\Psi^2\rangle (t) \sim t^{-3}e^{-\tilde{\Gamma} t} + \langle(\Psi^\dagger)^2(\Psi)^2\rangle_{\mathrm{st}}.
\end{equation}
The latter result is in agreement with the non-relativistic limit of the diagonal term in \eqref{eq:phi4atationaryphase}. Furthermore, we note that the time evolution in the ShGM generically shows also oscillating behaviour due to the existence of off-diagonal terms. This can be attributed to the relativistic nature of the ShGM, which allows the crossing symmetry between the matrix element of a generic scattering process. 

In the LLM the late-time behaviour of the powers $(\Psi^\dagger)^n\Psi^n$ has a ladder structure~\cite{DeNardis-15}: $n=2k-1$ and $n=2k$ share the same exponent for any positive integer $k$. Moreover $n$ and $N$ are related via $k$ as $N=k$; so we arrive at the result 
\begin{equation}
\label{LateTimeLLn}
\langle(\Psi^\dagger)^n\Psi^n\rangle (t) \sim t^{-3n/2}e^{-\tilde{\Gamma} t} + \langle(\Psi^\dagger)^n(\Psi)^n\rangle_{\mathrm{st}}. 
\end{equation}
We do not observe any oscillating contributions: this is related to the gapless non-relativistic dispersion relation, since the oscillation frequency is proportional to the particle mass. 

%%%%%%%%%%%%%%%%%%%%%%%%%%%%%%%%%
\subsection{Discussion of the late-time behaviour}
\label{Discussion}
%%%%%%%%%%%%%%%%%%%%%%%%%%%%%%%%%
Previous results~\cite{Kormos-13pra,DeNardisCaux14,DeNardis-14,DeNardis-15,GranetEssler21} for the dynamics in the LLM do not show the exponential decay \eqref{LateTimeLLn}. There are several remarks to be made regarding this:
\begin{enumerate}
\item The appearance of exponential behaviour in our approach is due to the existence of annihilation poles in the form factors. If considering a quench to a non-interacting model, as is the case in the hard-core limit~\cite{Kormos-13pra,DeNardisCaux14}, such annihilation poles are absent in the relevant local form factors.\footnote{In the same way, the absence of annihilation poles in the form factors of the energy operator in the Ising model leads to pure power-law decay~\cite{FiorettoMussardo10}.} In our results, taking the hard-core limit in (\ref{NRK}) leads to a vanishing pair amplitude $\tilde{K}$. Hence the quench becomes trivial, in the sense that the initial state is the ground state of the system and the model is not driven out of equilibrium. In turn, the relaxation rate $\tilde{\Gamma}$ goes to zero. (We note that in the small quench regime we are employing in the LLM the only quench that ensures the absence of exponential decay is the trivial one.).
\item Another point to stress is the role of the initial state. The BEC state considered previously~\cite{DeNardisCaux14,DeNardis-14,DeNardis-15,GranetEssler21} is by construction different to the initial state we are considering (\ref{InitialSt}). While the BEC state possesses a fixed particle number, our initial state is a superposition of states with different particle numbers. However, the thermodynamic properties of the overlaps with the stationary state agree, as seen from the fact that \eqref{gTBAthLL} agrees with the gTBA equation in Ref.~\cite{DeNardisCaux14}. Thus the two setups agree at least in the stationary regime.
\item Furthermore, quenches from the BEC state to an interacting LLM have been studied~\cite{DeNardis-14,DeNardis-15,GranetEssler21} by means of a particle-hole expansion around the saddle point within the quench action method. The difference to our result \eqref{LateTimeLLn} can be attributed to the fact that the used expansions are intrinsically different, as we show in \ref{Quench Action method for Sinh-Gordon model} for the ShGM. While we expanded in excitations above the ground state, the particle-hole expansion stays in the basin of attraction of the stationary state~\cite{CauxEssler13}. As a consequence, for instance off-diagonal terms are not included in the latter. Second, each term in the particle-hole expansion is regular, i.e., the singular parts of the form factors responsible for exponential decay are absent. 
\item On the other hand, the linked cluster expansion we carried out focused on the most singular contributions leading to exponential decay. We did not aim to address the power-law tails, which could be identified as contributions which are not resummed into exponentials. While we did not consistently consider such terms here, in principle they are contained in the linked cluster expansion. Thus it may be that at sufficiently late times additional terms showing the power-law decay have to be added to \eqref{LateTimeLLn}. 
\end{enumerate}
Further work related to the mismatch will be the subject of a forthcoming publication.

%%%%%%%%%%%%%%%%%%%%%%%%%%%%%%%%%
\section{Conclusion}
\label{Conclusion}
%%%%%%%%%%%%%%%%%%%%%%%%%%%%%%%%%
We have considered the time evolution of different operators (powers of the field operators and vertex operators) after a global quench in the ShGM. We assumed the system to be initialised in a boundary state of the form (\ref{InitialSt}). The following relaxation was described in both the linked cluster approach and quench action method and it turned to be dominated by exponentially decay. In addition, oscillations and power-law decay were obtained in the late time dynamics.

We were then able to map these results and the initial state in a consistent way to the LLM by the means of non-relativistic limit. We obtained the stationary state expectation values and relaxation dynamics. In contrast to previous results, we found exponential decay in the LLM. We discussed several aspects of this discrepancy, however, at the moment a full understanding of this mismatch is missing. Since it was only possible to describe the repulsive regime of LLM as the non-relativistic limit of ShGM, an interesting outlook would be the description of the attractive regime of LLM starting from the sine-Gordon theory, given its relation to the Kardar--Parisi--Zhang equation~\cite{Calabrese-14}.

%%%%%%%%%%%%%%%%%%%%%%%%%%%%%%%%%
\section*{Acknowledgements}
%%%%%%%%%%%%%%%%%%%%%%%%%%%%%%%%%
We thank Bruno Bertini, Pasquale Calabrese, Jean-S\'{e}bastien Caux, Fabian Essler, Giuseppe Mussardo and Jacopo De Nardis for very useful discussions. This work is part of the D-ITP consortium, a program of the Dutch Research Council (NWO) that is funded by the Dutch Ministry of Education, Culture and Science (OCW).

%%%%%%%%%%%%%%%%%%%%%%%%%%%%%%%%%
\appendix
%%%%%%%%%%%%%%%%%%%%%%%%%%%%%%%%%
\section{Form factors}
\label{Form Factors: axioms and properties}
%%%%%%%%%%%%%%%%%%%%%%%%%%%%%%%%%
We collect some results on the form factors in the ShGM~\cite{Mussardo10} and in the LLM~\cite{KorepinBogoliubovIzergin93}.

%%%%%%%%%%%%%%%%%%%%%%%%%%%%%%%%%
\subsection{Sinh-Gordon form factors}
\label{Sinh-Gordon Form Factors}
%%%%%%%%%%%%%%%%%%%%%%%%%%%%%%%%%
The form factors of an operator $O$ are defined as
\begin{equation}
\label{FFDef}
F^{O}_{N}(\theta_1,\dots,\theta_N) = \langle 0 |O| \theta_1,\dots,\theta_N\rangle
=\bra{0}OZ^\dagger(\theta_1)\ldots Z^\dagger(\theta_N)\ket{0},
\end{equation}
with the Faddeev--Zamolodchikov operators $Z^\dagger(\theta)$ creating the excitations. Given that we are making frequent use of them for states populated by particle pairs, we fix their notation as follows
\begin{equation}
\label{FFNot}
F^{O}_{2N}(-\theta_N,\dots,\theta_N) = \langle 0 |O| -\theta_N,-\theta_{N-1},\dots,-\theta_1,\theta_1,\dots,\theta_{N-1},\theta_N\rangle.
\end{equation}
In both cases, (\ref{FFDef}) and (\ref{FFNot}), the rapidities are ordered as
\begin{equation}
	\theta_1<\theta_2<\dots<\theta_{N-1}<\theta_N,
\end{equation}
in the second case the rapidities must be positive.

The form factors satisfy the following axioms (see, e.g., References~\cite{Mussardo10,Smirnov92book}):
\begin{itemize}
\item Analyticity: The form factors $F^{O}_{N}(\theta_1,\dots,\theta_N) $ are meromorphic functions in the physical strip $0\le\mathfrak{Im}\theta_N\le 2\pi$.
\item Scattering axiom:
\begin{eqnarray}
\label{ScattAx}
&& F^{O}_{N}(\theta_1,\dots,\theta_i,\theta_{i+1},\dots,\theta_N) \nonumber\\
&&\qquad\qquad= S(\theta_i - \theta_{i+1})F^{O}_{N}(\theta_1,\dots,\theta_{i+1},\theta_i,\dots,\theta_N).
\end{eqnarray}
\item Periodicity axiom:
\begin{equation}
\label{PeriodAx}
F^{O}_{N}(\theta_1+2\pi\ii,\theta_2,\dots,\theta_N) = l(O)F^{O}_{N}(\theta_2,\dots,\theta_N,\theta_1),
\end{equation}
where $l(O)$ represents the mutual semi-locality factor between the operator $O$ and the fundamental field ones. All operators considered in this paper are local, i.e., $l(O)=1$.
\item Lorentz transformation:
\begin{equation}
\label{LorentzTr}
F^{O}_{N}(\theta_1+\Lambda,\dots,\theta_N+\Lambda)=e^{s(O)\Lambda}F^{O}_{N}(\theta_1,\dots,\theta_N),
\end{equation}
where $s(O)$ is the Lorentz spin of the operator. All operators considered in this paper are spinless, thus $e^{s(O)}=1$.
\item Annihilation pole axiom:
\begin{eqnarray}
\label{AnnPol}
&&\mbox{Res}\left[F^{O}_{N+2}(\theta',\theta,\theta_1,\dots,\theta_N),\theta'=\theta+\ii\pi\right] \nonumber\\
&&\qquad = \ii\left(1-l(O)\prod^N_{k=1}S(\theta-\theta_k)\right)F_N^{O}(\theta_1,\dots,\theta_N) 
\end{eqnarray}
In particular, for the two-particle form factor of the vertex operator we get
\begin{equation}
\mbox{Res}\left[F^{\alpha}_{2}(\theta',\theta),\theta'=\theta+\ii\pi\right] = 0.
\end{equation}
\end{itemize}
Furthermore, one can relate incoming and outgoing scattering states via the crossing symmetry, for example
\begin{equation}
\label{CrossAx}
F^{O}_{N+1}(\xi +\ii\pi,\theta_1,\dots,\theta_N) = \langle \xi |O| \theta_1,\dots,\theta_N\rangle.
\end{equation}

The full expression of the form factor of the vertex operator was found in~\cite{KoubekMussardo93},
\begin{eqnarray}
F^{\alpha}_{N}(\theta_1,\dots,\theta_N) &=& \frac{\sin{\alpha\pi B}}{\sin{\alpha\pi}}\left(\frac{4\sin{\pi B}}{F_{\mathrm{min}}(\ii\pi)}\right)^{N/2}\nonumber\\
& &\qquad\qquad\times\mbox{det}M_N(\alpha)\prod_{j<l}^N\frac{F_{min}(\theta_j-\theta_l)}{x_j+x_l},\label{VOFF}
\end{eqnarray}
where $x_j=e^{\theta_j}$, $M_N$ is a matrix of symmetric polynomials $\sigma_k^{(N)}(x_1,\dots,x_N)$,
\begin{equation}
\label{Mmatrix}
[M_N(\alpha)]_{j,l} = \sigma_{2j-l}^{(N)}\frac{\sin{(j-l+\alpha)\pi B}}{\sin{\pi B}},
\end{equation}
and the minimal form factor satisfies
\begin{equation}
\label{MFFCond}
F_{\mathrm{min}}(\theta)F_\mathrm{min}(\ii\pi + \theta) = \frac{\sinh{\theta}}{\sinh{\theta}+\ii\sin{\pi B}}.
\end{equation}
Analytic expressions are available also for the form factors of the field powers~\cite{Kormos-10}, but we shall recall here only their most important property, i.e.,
\begin{equation}
\label{FFFO}
F_N^{:\phi^n:}(\theta_1,\dots,\theta_N) = 0\quad \mathrm{if}\;N<n,
\end{equation}
which directly follows from the normal ordering prescription.

Matrix elements containing incoming and outgoing particles can be decomposed into connected and disconnected parts~\cite{Smirnov92book}: We denote the sets of incoming and outgoing particles by $A$ and $B$ respectively. If $A$ and $B$ are decomposed into subsets, $A=A_1\cup A_2$, $B=B_1\cup B_2$, and we denote the product of scattering matrices required to reorder the particles by $S_{AA_1}$ (and similarly for the set $B$), i.e., $\ket{A}=S_{AA_1}\ket{A_2A_1}$, then 
\begin{equation}
\label{SmirDec}
\langle A|O|B\rangle = \sum_{A=A_1\cup A_2}\sum_{B=B_1\cup B_2}S_{A A_1}S_{B B_1}\langle A_1|O|B_1\rangle_\mathrm{conn}\langle A_2|B_2\rangle.
\end{equation}
The connected form factors can be extracted as in (\ref{FFconn}), while the second factors are simple scalar products in the Hilbert space of the asymptotic states. This property holds independently from the relativistic nature of the system.

%%%%%%%%%%%%%%%%%%%%%%%%%%%%%%%%%
\subsection{Lieb--Liniger form factors}
\label{Lieb-Liniger Form Factors}
%%%%%%%%%%%%%%%%%%%%%%%%%%%%%%%%%
In the LLM form factors must be defined in a more general fashion,
\begin{equation}
\label{FFDefLL}
\tilde F^{\tilde{O}}_{N+M}(p_1,\dots,p_N|q_1,\dots,q_M) = \langle  p_1,\dots,p_N|\tilde{O}| q_1,\dots,q_M\rangle
\end{equation}
This sharp distinction between in- and out- states is necessary since there is no crossing symmetry (\ref{CrossAx}) in the non-relativistic regime. In addition, the form factors do not depend on rapidities but rather momenta, analytic continuation and thus the periodicity axiom do not hold, and the annihilation pole axiom has to be rephrased as discussed in Reference~\cite{KorepinBogoliubovIzergin93}. It was shown~\cite{Kormos-10} that this definition can also be obtained by the means of non-relativistic limit.

A fundamental property of these form factors is that, for the operators $\left(\Psi^\dagger\right)^P\Psi^Q$, the only non-vanishing matrix elements are provided by the in-state containing $N+P$ particles and the out-state containing $N+Q$. Our specific case is characterised by having $P=Q=n$ which means that the off-diagonal matrix element are zero.

%%%%%%%%%%%%%%%%%%%%%%%%%%%%%%%%%
\section{Linked cluster expansion for sinh-Gordon model}
\label{Linked Cluster expansion for Sinh-Gordon model}
%%%%%%%%%%%%%%%%%%%%%%%%%%%%%%%%%
The linked cluster approach is developed by considering the expansion of the expectation values for small pair amplitudes $K(\theta,R)$. First, we discuss it for a general operator $O$ in the first two non-trivial contributions of lowest order. Then we are going to specify our computations for the different cases, i.e., $O=:\phi^{2n}:$ and $O=e^{\alpha g\phi}$.

Special care will be taken to isolate time invariant contributions from the result; we expect them to return the stationary values of the observables after the quench, while the remainder shall give us the late time corrections. Another time independent contribution, contained in the double pole extraction, comes from taking the rapidity derivative for $K$ functions; we are going to show that this one must vanish in our setup, but not in general.

%%%%%%%%%%%%%%%%%%%%%%%%%%%%%%%%%
\subsection{Linked cluster expansion for a generic operator} 
\label{Linked Cluster expansion for a generic operator}
%%%%%%%%%%%%%%%%%%%%%%%%%%%%%%%%%
An operator independent quantity is the quench partition function, namely the braket $\langle\psi|\psi\rangle$. In the small quench limit, this can be expanded as shown in (\ref{DenExp}). Moreover this expansions picks up the same diverging contributions in the infinite-volume limit $L\to\infty$; they are exactly the so-called linked clusters. We can compute the first contribution $Z_2$, which is useful for the rest of the chapter
\begin{equation}
	Z_2 = \int_{0}^{\infty}\frac{\dd\xi}{2\pi}\frac{\dd\theta}{2\pi}K^*(\xi)K(\theta)\langle-\xi,\xi|-\theta,\theta\rangle.
\end{equation}
As one can see, the scalar product yields the squared Dirac delta function $\delta^2(\xi-\theta)$. By regularising it with (\ref{LVDeltaReg}) we get
\begin{equation}
	\label{Z2}
	Z_2 = mL\int_{0}^{\infty}\frac{\dd\theta}{2\pi}|K(\theta)|^2\cosh{\theta}+\mathcal{O}\left(\frac{1}{L}\right),
\end{equation}
which is finite for a finite volume $L$. The information about the initial state is contained in the pair amplitude $K$.

Once we focus on the numerator of the expansion, for any operator $O$, we must at first define the lowest non-zero order in the expansion (\ref{NumExp}). In some cases, as for the vertex operator, it is just the one given by the post-quench ground state $|0\rangle$, i.e., the vacuum expectation value. In others, like any normal ordered power of the field operator, $:\phi^{2n}:$, the lowest order contribution is of order $n$. We label this quantity as $A$, e.g., $A=0$ for the vertex operator. This contribution then is particularly straightforward to write down 
\begin{eqnarray}
	\label{LCLO}
	\mathcal{C}^O(t) &=  \sum_{i,j=0}^A \frac{\delta_{j,A-i}}{i!j!}C_{2i,2j}^{O}(t) =\sum_{i=0}^A \frac{1}{i!(A-i)!}C_{2i,2(A-i)}^{O}(t)\\
	& = \sum_{i=0}^A\frac{1}{i!(A-i)!}\int_{0}^{\infty}\prod_{a=1}^i\frac{\dd\xi_a}{2\pi}K^*(\xi_a)\prod_{b=1}^{A-i}\frac{\dd\theta_b}{2\pi}K(\theta_b)\nonumber\\*
	&\qquad\qquad\times F_{2A}^{O}(-\xi_i+\ii\pi,-\xi_{i-1}+\ii\pi,\dots,\xi_i+\ii\pi,-\theta_{A-i},\dots,\theta_{A-i})\nonumber\\*
	&\qquad\qquad\times e^{2\ii mc^2t\left(\sum_{a=1}^{i}\cosh{\xi_a}-\sum_{b=1}^{A-i}\cosh{\theta_b}\right)}\nonumber
\end{eqnarray}
As one can see, no further simplification can be made, given the fact that the form factor does not possess annihilation poles and connected parts. The late-time behaviour can be extracted through a saddle-point approximation, but the more interesting contributions to isolate are the time independent, namely the stationary ones. In order to do that, we start from the following consideration: the full time dependence is contained in the exponential term; this is a constant if and only if $\sum_{a=0}^{i}\cosh{\xi_a}-\sum_{b=0}^{j}\cosh{\theta_b}=0$. This requires $i=j=A$, and then picks a region of integration where $\xi_a=\theta_a$. Once we restrict to this region, we get the following stationary contribution 
\begin{eqnarray}
	\label{LCstLO}
	&\mathcal{C}^O_\mathrm{st}=\frac{1}{A!^2}\int_{0}^{\infty}\prod_{a=1}^A\frac{\dd\theta_a}{2\pi}|K(\theta_a)|^2 F_{2A}^{O}(-\theta_A+\ii\pi,-\theta_{A-1}+\ii\pi,\dots,\theta_A).
\end{eqnarray}
We find agreement between this expression and the usual LeClair--Mussardo series (\ref{LMseries}), since here the form factor is indeed connected (even though we kept that implicit) and the first term is the leading order small quench expansion [as we already stated in (\ref{PartDensSol})] of the filling fraction. It is important to keep in mind that the same procedure can be carried out for any other order in the linked cluster expansion, provided that the disconnected parts and the annihilation poles of the form factor have been previously extracted. In principle, when we resum the expansion, stationary contributions should give the full LeClair--Mussardo series.

The following term in the expansion has order $A+1$ in $K$ for small quenches. It does not contain interesting contributions, as we are going to show. The Smirnov decomposition for any matrix element of the operator $O$ with $2A+2$ rapidities is the following
\begin{eqnarray}
	\label{SmirA1}
	&\langle-\xi_{i+1},\dots,\xi_{i+1}|O|-\theta_j,\dots,\theta_j\rangle\\
	&=\sum_{d,d'}F_{2A}^{O}(\dots,\xi_{d-1}+\ii\pi,\xi_{d+1}+\ii\pi,\dots,\theta_{d'-1},\theta_{d'+1},\dots)\delta\left(\xi_d-\theta_{d'}\right)\nonumber\\
	& +\sum_{d,d'}F_{2A}^{O}(\dots,-\xi_{d+1}+i\pi,-\xi_{d-1}+i\pi,\dots,-\theta_{d'+1},-\theta_{d'-1},\dots)\nonumber\\*
	& \qquad\times\delta\left(\xi_d-\theta_{d'}\right)+ F_{2A+2}^{O}(-\xi_{i+1}+\ii\pi,\dots,\xi_{i+1}+\ii\pi,-\theta_j,\dots,\theta_j),\nonumber
\end{eqnarray}
where still $i+j+1=A+1$, and the sums are over all possible contractions, i.e., $d=1,\ldots,i+1$, $d'=1,\ldots, j$. Furthermore, we have taken one particle more for the ingoing state, without any loss of generality. Together with the analogous term with one extra particle in the outgoing state, the final expression is obtained taking the real part. From here one can simply realise that these contributions have neither poles nor a diverging part in the infinite-volume limit. Moreover, if one can extract stationary contributions from the lowest order term [namely the terms (\ref{LCstLO})], then the same cannot be done for the next-to-lowest order, since the condition $i=j$ cannot be fulfilled any longer. Hence these contributions are only given by terms oscillating in time, vanishing in the steady state and unimportant for the relaxation processes.

In contrast to what was observed in other works~\cite{SE12,BSE14,CS17}, contributions coming from non-entangled particles with the same rapidity [i.e. second and third term on the right-hand side of Equation (\ref{SmirA1})] in the in- and out-state are present. This happens since we are considering operators whose $A$ is different from zero, namely the normal ordered field powers.

On the other hand, the terms of order $A+2$ possess a much richer structure. Here the Smirnov factorisation splits the contribution into three different parts: the fully disconnected term, the semi-connected and fully connected ones. The first term is shown to be proportional to (\ref{Z2}); the corresponding contribution in the expansion is
\begin{eqnarray}
	\label{DiscA2}
	&\sum_{i,j=0}^{A+2}\sum_{d=1}^{i}\sum_{d'=1}^{j}\frac{\delta_{j,A+2-i}}{i!j!}\int_{0}^{\infty}\prod_{a=1}^i\frac{\dd\xi_a}{2\pi}K^*(\xi_a)\prod_{b=1}^j\frac{\dd\theta_b}{2\pi}K(\theta_b)\nonumber\\*
	&\quad\times F_{2A}^{O}(\dots,-\xi_{d+1}+\ii\pi,-\xi_{d-1}+\ii\pi,\dots,\xi_{d-1}+\ii\pi,\xi_{d+1}+\ii\pi,\dots,\nonumber\\*
	&\qquad\qquad,-\theta_{d'+1},-\theta_{d'-1},\dots,\theta_{d'-1},\theta_{d'+1},\dots)\delta^2{(\theta_{d'}-\xi_d)}\nonumber\\*
	&\quad\times e^{2\ii mc^2t\left(\sum_{a=1}^{i}\cosh{\xi_a}-\sum_{b=1}^{j}\cosh{\theta_b}\right)}.
\end{eqnarray}
Regularising the squared Dirac delta and summing over the different combinations of the disconnected part, we get exactly $\mathcal{C}^{O}(t)Z_2$. In the expansion (\ref{LC2}), this term is cancelled by the expansion of the denominator. Since this term is diverging in the infinite volume limit, the final contribution at this order is independent of $L$. 

The semi-connected contribution contains as well a singularity
\begin{eqnarray}
	\label{SemiA2}
	&\sum_{i,j=0}^{A+2}\sum_{d=1}^{i}\sum_{d'=1}^{j}\frac{\delta_{j,A+2-i}}{i!j!}\int_{0}^{\infty}\prod_{a=1}^i\frac{\dd\xi_a}{2\pi}K^*(\xi_a)\prod_{b=1}^j\frac{\dd\theta_b}{2\pi}K(\theta_b)\\*
	&\quad\times F_{2A+2}^{O}(\dots,-\xi_{d+1}+\ii\pi+\ii\epsilon,-\xi_{d-1}+\ii\pi+\ii\epsilon,\dots,-\theta_{d'+1},-\theta_{d'-1},\dots)\nonumber\\*
	&\qquad\times \delta{(\theta_{d'}-\xi_d)} e^{2\ii mc^2t\left(\sum_{a=1}^{i}\cosh{\xi_a}-\sum_{b=1}^{j}\cosh{\theta_b}\right)}\nonumber\\
	&+\int_{0}^{\infty}\prod_{a=1}^i\frac{\dd\xi_a}{2\pi}K^*(\xi_a)\prod_{b=1}^j\frac{\dd\theta_b}{2\pi}K(\theta_b)\nonumber\\*
	&\quad\times F_{2A+2}^{O}(\dots,\xi_{d-1}+\ii\pi+\ii\epsilon,\xi_{d+1}+\ii\pi+\ii\epsilon,\dots,\theta_{d'-1},\theta_{d'+1},\dots)\nonumber\\*
	&\qquad\times \delta{(\theta_{d'}-\xi_d)} e^{2\ii mc^2t\left(\sum_{a=1}^{i}\cosh{\xi_a}-\sum_{b=1}^{j}\cosh{\theta_b}\right)}\nonumber
\end{eqnarray}
since the Dirac delta forces the form factor to take the value across the simple annihilation pole. These values are formally regularised by $\ii\epsilon$ and we can prove that, in this scheme, they cancel out with the boundary term in (\ref{SPTh}), when applied to the double pole of the connected contribution. But in order to do this, we need to discuss the fully connected contribution.

The last piece contains most of the relaxation dynamics features of the model. The form factor is expanded in the neighbourhood of annihilation poles as 
\begin{eqnarray}
	\label{DPFF}
	F_{2A+4}^{O}(-\xi_{i}+i\pi,\dots, \theta_{j}) \sim &\sum_{i,j=0}^{A+2}\sum_{d=1}^{i}\sum_{d'=1}^{j}\delta_{j,A+2-i}\frac{4}{\left(\xi_{d} - \theta_{d'}\right)^2}\nonumber\\*
	&\times F_{4A}^{\mathcal{O}}(-\xi_i+\ii\pi,\dots,\xi_i+\ii\pi,-\theta_j,\dots,\theta_j)\\
	&\qquad + \mbox{regular terms}\nonumber
\end{eqnarray}
As one can see, the remainder is the same contribution we got in (\ref{LCLO}). We are going to compute the integral by the means of (\ref{SPTh}). The derivative over the rapidity brings down one power in time, i.e., a linear term appears. The other contributions are vanishing in our quench setup. The fully connected part is the following
\begin{eqnarray}
	\label{ConnA2} &\sum_{i,j=0}^{A+2}\frac{\delta_{j,A+2-i}}{i!j!}\int_{0}^{\infty}\prod_{a=1}^i\frac{\dd\xi_a}{2\pi}K^*(\xi_a)\prod_{b=1}^j\frac{\dd\theta_b}{2\pi}K(\theta_b)\nonumber\\*
	&\qquad\times F_{2A+4}^{O}(-\xi_i+\ii\pi,\dots,\xi_i+\ii\pi,-\theta_j,\dots,\theta_j)\nonumber\\*
	&\qquad\times e^{2\ii mc^2t\left(\sum_{a=1}^{i}\cosh{\xi_a}-\sum_{b=1}^{j}\cosh{\theta_b}\right)}.
\end{eqnarray}
Extracting the double pole, we have to sum over the different ways to take it, with the result being
\begin{equation}
	\label{ConnA21}
	-\Gamma t\mbox{ }\mathcal{C}^O(t) + \mbox{regular terms} + \dots,
\end{equation}
where $\Gamma$ is the relaxation rate given in (\ref{Gamma}), and the regular terms give corrections to the late-time behaviour and stationary values.

When all the contributions are summed up, we get
\begin{eqnarray}
	\label{LCGO}
	\langle O(t)\rangle = (1-\Gamma t)\mathcal{C}^O(t) + \mbox{regular terms} + \mathcal{C}_\mathrm{st}^{O}+\mbox{corrections}.
\end{eqnarray}
This is as far as we get if we consider only the lowest orders; in order to go beyond we would need to take into account other features that show up at higher orders. A clever way to do this is using the representative state technique. Before doing so, we specify the linked cluster expansion to the operators we are interested in.

%%%%%%%%%%%%%%%%%%%%%%%%%%%%%%%%%
\subsection{Field powers expectation values}
\label{Field powers expectation values}
%%%%%%%%%%%%%%%%%%%%%%%%%%%%%%%%%
The case of field powers follows straightforwardly from the previous analysis. We recover the same structure of (\ref{LCGO}) and we can also specify the extra regular terms in it. For any power, we get
\begin{eqnarray}
	\label{LOFP}
	\mathcal{C}^{:\phi^{2n}:}(t) = &\sum_{i,j=0}^n\frac{\delta_{j,n-i}}{i!j!}\int_{0}^{\infty}\prod_{a=1}^i\frac{\dd\xi_a}{2\pi}K^*(\xi_a)\prod_{b=1}^j\frac{\dd\theta_b}{2\pi}K(\theta_b)\nonumber\\*
	&\qquad\times F_{2n}^{:\phi^{2n}:}(-\xi_i+\ii\pi,\dots,\xi_i+\ii\pi,-\theta_j,\dots,\theta_j)\nonumber\\*
	&\qquad\times e^{2\ii mc^2t\left(\sum_{a=0}^{i}\cosh{\xi_a}-\sum_{b=0}^{j}\cosh{\theta_b}\right)}.
\end{eqnarray}
For $n=1,2$ we get shorter expressions:
\begin{eqnarray}
	\label{LOFP1}
	\mathcal{C}^{:\phi^{2}:}(t) &= \mathfrak{Re}\int_0^{\infty}\frac{\dd\theta}{2\pi}K(\theta)F_2^{:\phi^2:}(-\theta,\theta)e^{-2\ii mc^2t\cosh{\theta}},\\
	\label{LOFP2}
	\mathcal{C}^{:\phi^{4}:}(t)&=\int_0^{\infty}\frac{\dd\xi}{2\pi}\frac{\dd\theta}{2\pi}K^*(\xi)K(\theta)F_4^{:\phi^4:}(-\xi+\ii\pi,\xi+\ii\pi,-\theta,\theta)\\*
	&\qquad\qquad\times e^{2\ii mc^2t(\cosh{\xi}-\cosh{\theta})}\nonumber\\
	&\quad+ \frac{1}{2}\mathfrak{Re}\int_0^{\infty}\frac{\dd\xi_1}{2\pi}\frac{\dd\xi_2}{2\pi}K^*(\xi_1)K^*(\xi_2)F_4^{:\phi^4:}(-\xi_2,-\xi_1,\xi_1,\xi_2)\nonumber\\*
	&\qquad\qquad\times e^{2\ii mc^2t(\cosh{\xi_1}+\cosh{\xi_2})}. \nonumber
\end{eqnarray}
The latter contains also stationary terms: we isolate them, since they are all contained inside the diagonal term, i.e., if $\xi=\theta$ we get 
\begin{equation}
	\label{phi4st}
	\mathcal{C}_\mathrm{st}^{:\phi^4:} = \int_0^{\infty}\frac{\dd\theta}{2\pi}|K(\theta)|^2F_\mathrm{4,conn}^{:\phi^4:}(-\theta+\ii\pi,\theta+\ii\pi,-\theta,\theta).
\end{equation}
The general stationary contribution occurs at the lowest order for even $n$. Its expression follows from (\ref{LCstLO}) with $A=n$ and the correct form factor. For odd $n$, the first stationary contribution has order $n+1$; for instance, for $n=1$
\begin{equation}
	\label{phi2st}
	\mathcal{C}_\mathrm{st}^{:\phi^2:}=\int_0^{\infty}\frac{\dd\theta}{2\pi}|K(\theta)|^2 F_\mathrm{4,conn}^{:\phi^2:}(-\theta+\ii\pi,\theta+\ii\pi,-\theta,\theta).
\end{equation}
The final result of the expansion is then, for these two operators
\begin{eqnarray}
	\label{LCphi2App}
	&\langle :\phi^2:\rangle (t) = 2(1-\Gamma t)\mathfrak{Re}\int_0^{\infty}\frac{\dd\theta}{2\pi}K(\theta)F_2^{:\phi^2:}(-\theta,\theta)e^{-2\ii mc^2t\cosh{\theta}}\\
	& \qquad\qquad\qquad+ \int_0^{\infty}\frac{\dd\theta}{2\pi}|K(\theta)|^2 F_\mathrm{4,conn}^{:\phi^2:}(-\theta+\ii\pi,\theta+\ii\pi,-\theta,\theta) \nonumber\\
	& \qquad\qquad\qquad+ \mbox{regular parts} + \mathcal{O}(K^4),\nonumber
\end{eqnarray}
\begin{eqnarray}
	\label{LCphi4App}
	&\langle :\phi^4:\rangle (t) =(1-\Gamma t)\int_0^{\infty}\frac{\dd\xi}{2\pi}\frac{\dd\theta}{2\pi}K^*(\xi)K(\theta)\\*
	&\qquad\qquad\qquad\times F_4^{:\phi^4:}(-\xi+\ii\pi,\xi+\ii\pi,-\theta,\theta) e^{2\ii mc^2t(\cosh{\xi}-\cosh{\theta})}\nonumber\\
	&\qquad + 2(1-\Gamma t)\mathfrak{Re}\int_0^{\infty}\frac{\dd\xi_1}{2\pi}\frac{\dd\xi_2}{2\pi}K^*(\xi_1)K^*(\xi_2) \nonumber\\*
	& \qquad\qquad\qquad\times F_4^{:\phi^4:}(-\xi_2,-\xi_1,\xi_1,\xi_2) e^{2\ii mc^2t(\cosh{\xi_1}+\cosh{\xi_2})} + \nonumber\\
	& \qquad+ \int_0^{\infty}\frac{\dd\theta}{2\pi}|K(\theta)|^2F_\mathrm{4,conn}^{:\phi^4:}(-\theta+\ii\pi,\theta+\ii\pi,-\theta,\theta) \nonumber\\
	& \qquad+ \mbox{regular parts} + \mathcal{O}(K^6). \nonumber
\end{eqnarray}

%%%%%%%%%%%%%%%%%%%%%%%%%%%%%%%%%
\subsection{Exponential operator expectation value} 
%%%%%%%%%%%%%%%%%%%%%%%%%%%%%%%%%
As anticipated before, the form factor of the exponential operator $e^{\alpha g\phi}$ has a simpler pole structure than the field powers. Moreover, the lowest order contribution is given by the ground state itself, i.e., $A=0$. We can then straightforwardly compute the first two orders
\begin{equation}
\label{C00exp}
C_{00}^{\alpha}(t) = \mathcal{G}(\alpha),
\end{equation}
\begin{equation}
\label{C02exp}
C_{02}^{\alpha}(t) + C_{20}^{\alpha}(t) = 2\,\mathfrak{Re}\int_0^{\infty}\frac{\dd\theta}{2\pi}K(\theta)F_2^\alpha(-\theta,\theta)e^{-2\ii mt\cosh{\theta}}.
\end{equation}
Where the vacuum expectation value of the exponential operator has been computed in~\cite{LukyanovZamolodchikov97}
\begin{eqnarray}
\label{expVEV}
&\mathcal{G}(\alpha) = \left[\frac{m\Gamma\left(\frac{2-B}{4}\right)\Gamma\left(\frac{4+B}{2}\right)}{4\sqrt{\pi}}\right]^{2\alpha^2}\\*
&\qquad\times\exp\left\{\int_0^{\infty}\frac{\dd t}{t}\left[\frac{\sin^2(2\alpha gt)}{2\sinh{(g^2t)\sinh{t}\cosh{(1+g^2)t}}} - 2\alpha^2e^{-2t}\right]\right\}, \nonumber
\end{eqnarray}
where the $\Gamma(z)$ denotes the Euler gamma function.

The diagonal second order term of the linked cluster expansion is again a fully connected one, for the locality of the exponential operator
\begin{eqnarray}
\label{C22exp}
C_{22}^{\alpha}(t) = &\int_0^{\infty}\frac{\dd\xi}{2\pi}\frac{\dd\theta}{2\pi}K^*(\xi)K(\theta)\\*
&\qquad\qquad\times F_4^\alpha(-\xi+\ii\pi,\xi+\ii\pi,-\theta,\theta) e^{2\ii mc^2t(\cosh{\xi}-\cosh{\theta})}.\nonumber
\end{eqnarray}
Then we can extract the stationary contribution from it,
\begin{eqnarray}
\label{C22exp'}
&C_\mathrm{22,st}^{\alpha} = \int_0^{\infty}\frac{\dd\xi}{2\pi}|K(\xi)|^2F_\mathrm{4,conn}^\alpha(-\xi+\ii\pi,\xi+\ii\pi,-\xi,\xi),
\end{eqnarray}
which has the same structure as (\ref{phi2st}) and (\ref{phi4st}). Please note that the form factor in (\ref{C22exp}) does not contain any singularity and we do not need to further regularise it. The remainder of the second-order contribution is an oscillating factor
\begin{eqnarray}
\label{C04and40}
&C_{04}^{\alpha}(t) + C_{40}^{\alpha}(t) \\
& = 2\,\mathfrak{Re}\int_0^{\infty}\frac{\dd\theta_1}{2\pi}\frac{\dd\theta_2}{2\pi}K(\theta_1)K(\theta_2)F_4^\alpha(-\theta_2,-\theta_1,\theta_1,\theta_2)e^{-2\ii mc^2t(\cosh{\theta_1}+\cosh{\theta_2})}.\nonumber
\end{eqnarray}
Our final result is then
\begin{eqnarray}
\label{LCexpApp}
&\langle e^{\alpha g\phi}\rangle (t) = \mathcal{G}(\alpha) \\
& \qquad+  \int_0^{\infty}\frac{\dd\xi}{2\pi}|K(\xi)|^2F_\mathrm{4,conn}^\alpha(-\xi+\ii\pi,\xi+\ii\pi,-\xi,\xi) \nonumber\\
& \qquad+ 2\,\mathfrak{Re}\int_0^{\infty}\frac{\dd\theta}{2\pi}K(\theta)F_2^\alpha(-\theta,\theta)e^{-2\ii mc^2t\cosh{\theta}} \nonumber\\
& \qquad+ 2\,\mathfrak{Re}\int_0^{\infty}\frac{\dd\theta_1}{2\pi}\frac{\dd\theta_2}{2\pi}K(\theta_1)K(\theta_2)F_4^\alpha(-\theta_2,\dots,\theta_2)e^{-2\ii mc^2t(\cosh{\theta_1}+\cosh{\theta_2})} \nonumber\\
& \qquad+ \mbox{regular parts} + \mathcal{O}(K^3).\nonumber
\end{eqnarray}
The first line contains the stationary contributions at this level of the linked cluster expansion; the double poles contributions are present at higher orders, since the locality of the operator, and they contribute to the exponential decay (\ref{LCexp}). 

A further check for these stationary values might be given in terms of Negro--Smirnov formula, where the ratio
\begin{eqnarray}
\label{NSForm}
&\frac{\langle e^{(\alpha+1)g\phi}\rangle_\mathrm{st}}{\langle e^{\alpha g\phi}\rangle_\mathrm{st}}\\
&\quad= \frac{\mathcal{G}(\alpha+1) +  \int_0^{\infty}\frac{\dd\xi}{2\pi}|K(\xi)|^2F_\mathrm{4,conn}^{\alpha+1}(-\xi+\ii\pi,\xi+\ii\pi,-\xi,\xi)}{\mathcal{G}(\alpha) +  \int_0^{\infty}\frac{\dd\xi}{2\pi}|K(\xi)|^2F_\mathrm{4,conn}^\alpha(-\xi+\ii\pi,\xi+\ii\pi,-\xi,\xi)}\nonumber
\end{eqnarray}
is considered. In the linked cluster formalism, it can be expanded in the following way
\begin{eqnarray}
\label{LCNS}
&\frac{\langle e^{(\alpha+1)g\phi}\rangle_\mathrm{st}}{\langle e^{\alpha g\phi}\rangle_\mathrm{st}}\\
&\quad= \frac{\mathcal{G}(\alpha+1)}{\mathcal{G}(\alpha)}\left[1+\int_0^{\infty}\frac{\dd\xi}{2\pi}|K(\xi)|^2\frac{F_\mathrm{4,conn}^{\alpha+1}(-\xi+\ii\pi,\xi+\ii\pi,-\xi,\xi)}{\mathcal{G}(\alpha+1)}\right.\nonumber\\
&\qquad\left.-\int_0^{\infty}\frac{\dd\xi}{2\pi}|K(\xi)|^2\frac{F_\mathrm{4,conn}^{\alpha}(-\xi+\ii\pi,\xi+\ii\pi,-\xi,\xi)}{\mathcal{G}(\alpha)}\right]. \nonumber
\end{eqnarray}
On the other hand, the Negro--Smirnov formula \eqref{NSform} yields, in the small quench regime,
\begin{eqnarray}
\label{NSexp}
&\frac{\langle e^{(\alpha+1)g\phi}\rangle_\mathrm{st}}{\langle e^{\alpha g\phi}\rangle_\mathrm{st}} = \frac{\mathcal{G}(\alpha+1)}{\mathcal{G}(\alpha)}\left[1 + \frac{2\sin{(\pi B(2\alpha+1))}}{\pi}\int_{-\infty}^{\infty}\dd\theta|K(\theta)|^2\right], \\
&p_\alpha(\theta) = e^{-\theta} + \mathcal{O}(K^2).
\end{eqnarray}
The two results agree since the connected form factors are constant and their difference gives the correct prefactor in (\ref{NSexp}).

%%%%%%%%%%%%%%%%%%%%%%%%%%%%%%%%%
\section{Quench action method for sinh-Gordon model}
\label{Quench Action method for Sinh-Gordon model}
%%%%%%%%%%%%%%%%%%%%%%%%%%%%%%%%%
In the quench action approach we compute the time evolution using (\ref{QAM}) and (\ref{RhoNorm}), using a representative eigenstate $\rho$,
\begin{equation}
\label{RhoDef}
|\rho\rangle = \lim_{N\to\infty}|-\rho_N,\dots,-\rho_1,\rho_1,\dots,\rho_N\rangle.
\end{equation}
As before we will take advantage of the large but finite-volume framework.

%%%%%%%%%%%%%%%%%%%%%%%%%%%%%%%%%
\subsection{Field powers expectation values}\label{QA1}
%%%%%%%%%%%%%%%%%%%%%%%%%%%%%%%%%
The computation for both cases ($:\phi^2:$ and $:\phi^4:$) is very similar, hence we present the full computation for the first case only. The argument can be easily generalised to higher powers and the vertex operators. The starting point is the formula (\ref{QAM})
\begin{equation}
\label{QAMphi2App}
\langle :\phi^2: \rangle (t) = \frac{1}{2}\lim_{L\to\infty}\left[\frac{\langle\psi|:\phi^2:(t) |\rho\rangle}{\langle\psi|\rho\rangle} + \frac{\langle\rho|:\phi^2:(t) |\psi\rangle}{\langle\rho|\psi\rangle}\right].
\end{equation}
The numerator can be expanded in the following way
\begin{eqnarray}
\label{Numphi2}
&\langle\psi|:\phi^2:(t)|\rho\rangle= \sum_{0<I_1<\dots<I_M}\mathcal{N}_{M}\prod_{j=1}^M K^*(\xi_j)e^{2\ii mc^2t\cosh{\xi_j}}\\*
&\qquad\qquad\qquad\times\prod_{l=1}^N e^{-2\ii mc^2t\cosh{\rho_l}}
\bra{-\xi_M,\ldots,\xi_M}:\phi^2:\ket{-\rho_N,\ldots,\rho_N},\nonumber
\end{eqnarray}
while the denominator becomes
\begin{equation}
\label{Denphi2}
\langle\psi|\rho\rangle = (mL)^N\mathcal{N}_N\prod_{j=1}^N K^*(\rho_j)\cosh{\rho_j} .
\end{equation}
The main task here is extracting the dominant time dependence, which clearly originates from (\ref{Numphi2}). From the linked cluster approach we know that the time dependence is mostly governed by the pole structure of the form factors. Hence for the case of $:\phi^2:$, the dominant contribution can be extracted from the $M=N+1$ term with the resulting most singular part given by
\begin{eqnarray}
\label{HOFF}
&\bra{-\xi_{N+1},\ldots,\xi_{N+1}}:\phi^2:\ket{-\rho_N,\ldots,\rho_N}\sim \sum_{a=1}^{N+1}F_2^{:\phi^2:}(-\xi_a,\xi_a)\\* 
&\qquad\qquad\qquad\qquad\times\prod_{b=1}^{N} \left[\frac{4}{\left(\xi_{\sigma_a(b)} - \rho_b\right)^2} + \delta^2(\xi_{\sigma_a(b)} - \rho_b)\right],\nonumber
\end{eqnarray}
where $\sigma_a(b)$ gives the label $b$ if $b<a$ and $(b+1)$ if $b>a$. 
Also, we have taken into account the disconnected parts of the form factor itself and the result factorises in the previous way.
The derivation of this formula is properly carried out in the next section.

We follow then with the same arguments of~\cite{BSE14}, except for a slight but necessary change: since the operator considered there mixes two different sectors in the finite-volume regime, the annihilation poles of the in-state are separated from those given by the out-state. In our case this is not true, but we can overcome this by simply considering the formula (\ref{HOFF}) and the relative finite-volume regularisation of the squared Dirac delta. As for the linked cluster expansion, the semi-connected pieces are not taken into account, since that they are cancelled by the singular part of the respective connected parts.

Then we can straightforwardly obtain the result
\begin{eqnarray}
\label{Numphi2'}
&\langle\psi| :\phi^2: (t) | \rho\rangle = \sum_{0<I_1<\dots<I_{N}}\mathcal{N}_{N}\prod_{j=1}^{N}K^*(\xi_j)e^{-2\ii mc^2t\cosh{\xi_j}}\\*
&\qquad\qquad\qquad\times\sum_{a=1}^{N}F_2^{:\phi^2:}(-\xi_a,\xi_a)\prod_{b=1}^{N} \left[\frac{4}{\left(\xi_{\sigma_a(b)} - \rho_b\right)^2} + \delta^2(\xi_{\sigma_a(b)} - \rho_b)\right],\nonumber
\end{eqnarray}
where the sum becomes an integral up to a vanishing contribution in the thermodynamic limit. After these considerations, poles can be safely picked from the most singular term and the delta contributions can be computed with (\ref{HOFF}), allowing us to arrive at 
\begin{eqnarray}
\label{QAMphi2'}
\frac{\langle\psi|:\phi^2:(t) |\rho\rangle}{\langle\psi|\rho\rangle} &= 2\int_0^{\infty}\frac{\dd\xi}{2\pi} K^*(\xi)F_2^{:\phi^2:}(-\xi,\xi)e^{-2\ii mc^2t\cosh{\xi}}\\*
&\qquad\qquad\qquad\times\lim_{N\to\infty}\prod_{j=1}^{N}\left(1-4mc^2\frac{t}{L}\tanh{\rho_j}\right).\nonumber
\end{eqnarray}
The integrand is the remainder after we take the kinetic poles and divide everything by the denominator.
Note that we are still retaining only the leading behaviour in $L$ of terms like $\rho$ and $\mathcal{N}$, which leads us to the result (\ref{QAMphi2'}).

Then we can take the thermodynamic limit $N\to\infty$ and $\frac{N}{L}\to1$, recovering this result
\begin{equation}
\label{QAMphi2res}
\langle:\phi^2: \rangle (t) = 2\,\mathfrak{Re}\int_0^{\infty}\frac{\dd\xi}{2\pi} K^*(\xi)F_2^{:\phi^2:}(-\xi,\xi)e^{2\ii mc^2t\cosh{\xi}}\, e^{-\Gamma t},
\end{equation}
where the relaxation rate $\Gamma$ is defined in \eqref{Gamma}.

In case of the operator $:\phi^4:$ or higher powers the same computations holds, provided we take care of the expression of the most singular part of the form factor and we identify correctly the dominant pole contribution. For a general field power $:\phi^{2n}:$, the dominant pole contribution in the braket $\langle\psi|:\phi^{2n}: |\rho\rangle$ is given by the sum over the terms
\begin{eqnarray}
N = M + n, M + n - 2, \dots, M+1,\quad n\;\mathrm{odd},\\
N = M + n, M + n - 2, \dots, M,\quad n\;\mathrm{even}.
\end{eqnarray}
The form factor in the vicinity of the annihilation poles can be expressed as
\begin{eqnarray}
\label{HOFFn}
&\langle-\xi_{N},\dots,\xi_N|:\phi^{2n}:|-\rho_{N-n},\dots, \rho_{N-n}\rangle\\
&\quad\sim \sum_{a_1,\dots,a_n=1}^{N}F_{2n}^{:\phi^{2n}:}(-\xi_{a_n},\dots,\xi_{a_n})\prod_{b=1}^{N-n} \left[\frac{4}{\left(\xi_{\sigma_a(b)} - \rho_b\right)^2} + \delta^2(\xi_{\sigma_a(b)} - \rho_b)\right].\nonumber
\end{eqnarray}
By defining again the quantity $\mathcal{C}^{\mathcal{\mathcal{O}}}(t)$ as we did for (\ref{LCLO}), we arrive at the following general result for the quench action method
\begin{equation}
\label{QAMn}
\langle:\phi^{2n}: \rangle(t) = \mathcal{C}_n(t)e^{-\Gamma t} + \langle:\phi^{2n}: \rangle_\mathrm{st}.
\end{equation}
The stationary expectation values of our operators cannot be determined in this framework. Assuming that is it sufficient to take the diagonal contributions $N=M$ and then apply the saddle point approximation, we would get
\begin{equation}
\label{stationaryEV}
\langle:\phi^{2n}: \rangle_\mathrm{st} = \lim_{L\to\infty}\langle\rho|:\phi^{2n}:|\rho\rangle,
\end{equation}
which was obtained on general ground in~\cite{CauxEssler13}.

Note also that the results (\ref{HOFFn}) and (\ref{QAMn}) can be applied to the exponential operator as well, provided that we identify $n=1$ [for the reason that (\ref{AnnPol}) holds] and
\begin{equation}
\label{Cexp}
\mathcal{C}^{\alpha}(t) =  C^{\alpha}_{20}(t) + C^{\alpha}_{02}(t).
\end{equation}
Hence we would analogously get
\begin{equation}
\label{QAMexpApp}
\langle e^{\alpha\phi} \rangle(t) = \mathcal{C}^{\alpha}(t)e^{-\Gamma t} + \langle e^{\alpha\phi} \rangle_\mathrm{st}.
\end{equation}
We can see that the vacuum expectation value contributes only to the stationary value, i.e., all the post-quench dynamics is generated by particle-pair excitations as expected.

%%%%%%%%%%%%%%%%%%%%%%%%%%%%%%%%%
\subsection{Proof of (\ref{HOFFn})}
\label{Proof 2}
%%%%%%%%%%%%%%%%%%%%%%%%%%%%%%%%%
Smirnov's decomposition formula (\ref{SmirDec}) states that form factors can be split into a sum of connected and disconnected ones: the first contain regularised annihilation poles, the second (squared) Dirac delta functions. As we show in \ref{Regularisation scheme}, the large but finite volume theory provides a finite result for both of them. Considering only the most diverging part of the form factors, we know that~\cite{BSE14} 
\begin{eqnarray}
\label{MDFF}
&F^{O}_{2(N+M)}(-\xi_N+\ii\pi,\dots,\xi_N+\ii\pi,-\theta_M,\dots,\theta_M)\\
&\quad\sim \sum_{i=0}^{|N-M|}F^{O}_{2|N-M|}(-\xi_i+\ii\pi,\dots,\theta_{|N-M|-i})\prod_{b=1}^M\frac{4}{(\xi_{\sigma_i(b)}-\theta_{\sigma_{|N-M|-i}(b)})^2},\nonumber
\end{eqnarray}
where the label $\sigma_i(b)$ gives back the $i$ indices that are not in the argument of the form factor. Now if we plug in the squared delta regularisation (\ref{LVDeltaReg}) we see that any term related to the difference between two rapidities appears as the double pole or as the Dirac delta. Hence, we can write this sum over all these strings as the product of the sum of the two divergent contributions for any of these differences and arrive at (\ref{HOFFn}).

This result is justified in light of the fact that we are considering only the most diverging pieces of the form factors, which can be factored out as (\ref{MDFF}). Also, contributions from the rearranging of the rapidities in the Smirnov formula are not taken into account.

%%%%%%%%%%%%%%%%%%%%%%%%%%%%%%%%%
\subsection{Particle-hole contributions}\label{QA2}
%%%%%%%%%%%%%%%%%%%%%%%%%%%%%%%%%
The quench action method provides also another way of computing the time evolution after a quantum quench. It can be straightforwardly derived from (\ref{QAM}) as done in~\cite{CauxEssler13} by introducing small excitations $\bold{e}$ over the steady state $\rho$,
\begin{eqnarray}
&\lim_{L\to\infty}\langle:\phi^{2n}:\rangle_L (t) \\
&\quad= \mathfrak{Re}\sum_{M=0}^{\infty}\int_{0}^{\infty}\prod_{j=1}^M\frac{\dd\theta_j}{2\pi}\rho(\theta_j)\frac{\dd\xi_j}{2\pi}\rho^{\mathrm{(h)}}(\xi_j) e^{-\delta s_{\bold{e}}}\langle \rho|:\phi^{2n}: (t)|\rho', \bold{e}_M \rangle_L,\nonumber
\end{eqnarray}
where, according to the Bethe ansatz, the excitation $\xi_j$ over the stationary state is obtained by changing one quantum number (which we will call hole) into another allowed one $\theta_j$ (named particle). In the large-volume regime, sums over holes and particles can be approximated by integrals. The remaining rapidities $\rho'$ are only weakly dependent on the particle-hole excitation [namely only for terms $\mathcal{O}(1/L)$], hence we can consider them equal to the unperturbed ones.

Now we focus on the matrix element: given that both the in- and the out- states are eigenstates of the Hamiltonian, the formal time dependance is rather simple,
\begin{equation}
\langle \rho|:\phi^{2n}: (t)|\rho', \bold{e}_M \rangle_L = \langle \rho|:\phi^{2n}:|\rho', \bold{e}_M \rangle_L e^{2\ii mc^2t\sum_{j=0}^M(\cosh{\theta_j}-\cosh{\xi_j})}.
\end{equation}
Moreover since we know the explicit form of the initial state, we can also write down the overlap exponential as
\begin{eqnarray}
	\label{Overlap}
&e^{-\delta s_{\bold{e}}} = \frac{\langle\psi|\rho', \bold{e}_M \rangle_L}{\langle\psi|\rho\rangle_L}, \\
&\langle\psi|\rho\rangle_L = (mL)^N \prod_{j=1}^N K(\rho_j)\cosh{\rho_j}, \\
&\langle\psi|\rho',\bold{e}_M\rangle_L = (mL)^N \prod_{j=1}^{N-M} K(\rho'_j)\cosh{\rho'_j}\prod_{i=1}^{M} K(\theta_i)\cosh{\theta_i}.
\end{eqnarray}
Wrapping everything up we get
\begin{equation}
e^{-\delta s_{\bold{e}}} = F_L(\rho,\rho')\prod_{j=1}^M\frac{K(\theta_j)\cosh{\theta_j}}{K(\xi_j)\cosh{\xi_j}},
\end{equation}
where the function $F_L$ contains the sub-leading corrections in the thermodynamic limit, $\lim_{L\to\infty}F_L=1$. 

Thus in the finite volume the matrix element $\bra{\rho}:\phi^2:\ket{\rho', \bold{e}_M}_L$ contains only the connected contributions since the two states differ by $M$ particles.
Moreover, the excited particles cannot take the value of the holes otherwise we would be double counting the zero-order terms.
Hence, one can proceed by taking the thermodynamic limit, considering only connected form factors in order to rule out these contributions.

Now by truncating the sum at the first correction, we can study the contributions of the state of one particle-hole excitation,
\begin{eqnarray}
&\lim_{L\to\infty}\langle:\phi^{2n}:\rangle_L (t) =\langle:\phi^{2n}:\rangle_\mathrm{st} \\
&\qquad+ \mathfrak{Re}\int_{0}^{\infty}\frac{\dd\theta}{2\pi}\frac{\dd\xi}{2\pi}\rho(\theta)\rho^{\mathrm{(h)}}(\xi) e^{-\delta s(\xi, \theta)}\langle \rho|:\phi^{2n}: (t)|\rho' (\xi, \theta) \rangle_L.\nonumber
\end{eqnarray}
Plugging in the previous expression with $M=1$, we get the correction 
\begin{eqnarray}
	\label{FOPH2n}
&\int_{0}^{\infty}\frac{\dd\theta}{2\pi}\frac{\dd\xi}{2\pi}\rho(\theta)\rho^{\mathrm{(h)}}(\xi)F_L(\xi,\theta)\frac{K(\theta)\cosh{\theta}}{K(\xi)\cosh{\xi}}\\*
&\qquad\qquad\qquad\times\langle\rho|:\phi^{2n}:|\rho' (\xi, \theta) \rangle_Le^{2\ii mc^2t(\cosh{\theta}-\cosh{\xi})}.\nonumber
\end{eqnarray}
Now the matrix element $\langle\rho|:\phi^{2n}:|\rho' (\xi, \theta) \rangle$ can be expressed as a series of ShGM connected form factors~\cite{CuberoPanfil20} in the infinite volume (for $L$ large enough). Given that (\ref{FOPH2n}) is continuous in the whole integration region, we can safely take the saddle-point approximation for large $t$, which gives the predicted $t^{-3}$ power law for any value of $n$~\cite{DeNardis-15}. Higher orders in $M$ correspond to sub-leading contributions in the late-time expansion.

The previous result is in stark contrast with those obtained for small quenches in the quench action approach (\ref{QAMn}). Not only the exponential term is absent, but also the explicit dependence between the power law exponent and the field power $n$ is missing. The reason for this is that the two expansions are incompatible: in the particle-hole expansion we are considering small values of $e^{-\delta s}$, which is independent of the magnitude of $K$ (\ref{Overlap}). On the other hand, $K$ is assumed to be small in the derivation of (\ref{QAMn}). From a physical perspective, in one case we expand in excitations over the ground state,  while in the other we are perturbing around the stationary state. The latter implies that the expansion in particle-hole pairs is only suitable at sufficiently late times, when exponential terms are already suppressed.

%%%%%%%%%%%%%%%%%%%%%%%%%%%%%%%%%
\section{Sinh-Gordon model in finite volume}
\label{Sinh-Gordon model in finite volume}
%%%%%%%%%%%%%%%%%%%%%%%%%%%%%%%%%
In this appendix we review the ShGM in a finite volume~\cite{PozsgayTakacs08-1,PozsgayTakacs08-2,KormosPozsgay10}, which provides a natural way to regularise the theory.

%%%%%%%%%%%%%%%%%%%%%%%%%%%%%%%%%
\subsection{Finite-volume theory}
%%%%%%%%%%%%%%%%%%%%%%%%%%%%%%%%%
The description of an integrable quantum field theory in a finite volume $L$ can be summarised in the following steps:
\begin{enumerate}
\item Build up a basis of energy eigenstates in the finite volume.
\item Define the form factors in this basis.
\item Choose the normalisation for these states in the Hilbert space.
\end{enumerate}
Application of the Bethe ansatz takes care of the first requirement; starting from the Bethe--Yang equations (\ref{BAShG}) we are able to introduce the following quantities:
\begin{eqnarray}
\label{FVDens}
&\mathcal{J}_{j,l}(\theta_1,\dots,\theta_N) = 2\pi\partial_j I_l = mL\cosh{\theta_l}\,\delta_{j,l} + \varphi(\theta_j - \theta_l), \\
&\rho_N(\theta_1,\dots,\theta_N) = \mbox{det}\mathcal{J}_{j,l}(\theta_1,\dots,\theta_N).
\end{eqnarray}
As we are not interested in the full energy spectrum at finite volume, we only need few results on the Jacobian matrix and its determinant generated by the mapping  (\ref{BAShG}) between rapidity variables $\theta_j$ and Bethe (half-)integers $I_j$. By the means of the latter, we are able to properly characterise any finite-volume state $|I_1,\dots,I_N\rangle_L$; the same notation we use for the infinite-volume states is valid as well for these ones. The initial states are then properly defined in the finite volume with the help of the Bethe ansatz formalism: 
\begin{eqnarray}
\label{FVInitialSt}
|\psi\rangle_L &=& \sum_{N=0}^{\infty}\sum_{0<I_1<\dots<I_N}\mathcal{N}_{2N}(\theta_1,\dots,\theta_N)\mathcal{K}_{2N}(\theta_1,\dots,\theta_N)\nonumber\\*
& &\qquad\qquad\qquad\times |-I_N,-I_{N-1},\dots,I_{N-1},I_N\rangle_L.
\end{eqnarray}
Here the normalisation and pair amplitude are given by 
\begin{eqnarray}
\label{NDef}
\mathcal{N}_{2N}(\theta_1,\dots,\theta_N)&=&\frac{\sqrt{\rho_{2N}(-\theta_N,-\theta_{N-1},\dots,\theta_N)}}{\rho_N(\theta_1,\dots,\theta_N)},\\
\label{KDef}
\mathcal{K}_{2N}(\theta_1,\dots,\theta_N)&=&K(\theta_1)\dots K(\theta_N).
\end{eqnarray}
Moreover, in the large but finite volume, the $\rho_N$ functions (which can be regarded as the Jacobian of the change of basis) have a simple shape, up to small corrections
\begin{equation}
\label{LVDens}
\rho_{N}(\theta_1,\dots,\theta_N) = (mL)^N\prod_{k=1}^N\cosh{\theta_k}\left[1+\mathcal{O}\left(\frac{1}{L}\right)\right].
\end{equation}
Thus the dominant contributions in the determinant are given by the diagonal ones. Furthermore, the volume dependence in the normalisation (\ref{NDef}) drops out.

When the representation (\ref{FVInitialSt}) is employed in order to compute expectation values in the large-volume regime, one subtlety must be considered: since the discrete sums are turned into integrals, the restriction over identical particles (given by the fermionic nature of the particles themselves) is loosened, which means that those states (i.e., those which contain two equal particles) also contribute to the final result. In principle we would need to subtract these contributions any time one considers integrals that overlap those regions; however they only show up in orders higher than those we consider.

Matrix elements are mapped into the new basis counterparts via (\ref{FVDens}), up to corrections $\mathcal{O}(e^{-mL})$, i.e.,
\begin{equation}
\label{FVFF}
\langle I_1,\dots,I_N|O|J_1,\dots,J_M\rangle_L\! = \!\frac{F^{O}_{N+M}(\xi_1+\ii\pi,\dots,\xi_N+\ii\pi,\theta_1,\dots,\theta_M)}{\sqrt{\rho_N(\xi_1,\dots,\xi_N)}\sqrt{\rho_M(\theta_1,\dots,\theta_M)}}.
\end{equation}
The normalisation of the states is chosen to be
\begin{equation}
\label{FVStNorm}
\langle I_1,\dots,I_N|J_1,\dots,J_M\rangle_L = \delta_{N,M}\delta_{I_1,J_1}\dots\delta_{I_N,J_M}.
\end{equation}

%%%%%%%%%%%%%%%%%%%%%%%%%%%%%%%%%
\subsection{Regularisation scheme}
\label{Regularisation scheme}
%%%%%%%%%%%%%%%%%%%%%%%%%%%%%%%%%
As for the infinite volume, form factors contain second order poles according to the annihilation pole axiom. They are also picked in the sum when the any of the entering and outgoing integers are equal, hence the finite volume does not seem to provide a natural way to regularise them. There are also other kind of singularities, arising from the very nature of the particle states; they are made of entangled parity-invariant pairs of fermions and it makes those states non-normalised, for their norm contains squared Dirac delta functions.

In the finite volume the states containing $N$ different particle pairs can be safely normalised by the means of the following formula
\begin{equation}
\label{NDeltaReg}
\prod_{j=1}^N\delta^2(\xi_j - \theta_j)= \frac{\rho_{2N}(-\xi_N,-\xi_{N-1}\dots,\xi_N)}{\rho_{N}(\xi_1,\dots,\xi_N)}
\prod_{j=1}^N\delta(\xi_j - \theta_j).
\end{equation}
The derivation of this formula will be presented below. So the regularisation of the squared Dirac delta functions depends in principle from the number of particles during the scattering process itself, i.e.,
\begin{eqnarray}
\label{DeltaReg}
\delta^2(\xi_j - \theta_j) &=& \frac{\rho_{2N}(-\xi_N,\dots,\xi_N)}{\rho_{2(N-1)}(-\xi_N,\dots,-\xi_{j+1},-\xi_{j+1},\dots,\xi_{j-1},\xi_{j+1},\dots,\xi_N)}\nonumber\\
& &\qquad\times \frac{\rho_{N-1}(\xi_{1},\dots,\xi_{j-1},\xi_{j+1},\dots,\xi_N)}{\rho_{N}(\xi_1,\dots,\xi_N)}\delta(\xi_j-\theta_j)\\
&=&mL\cosh{\xi_j}\delta(\xi_j-\theta_j),
\label{LVDeltaReg}
\end{eqnarray}
where in the last line we have explicitly performed the large-volume limit. This expression can be checked explicitly for the Ising model.

As we said before the finite-volume theory for the operators we are going to study does not give us a straightforward regularisation of the annihilation poles.
However in the thermodynamic limit sums are turned into integrals and, posing (\ref{DeltaReg}), the overall expression is finite, knowing that the pole sitting on the integration interval can be extracted by the means of Sokhotsky-Plemelj theorem
\begin{equation}
\label{SPTh}
\lim_{\epsilon\to0^+}\int_{0}^{\infty}\frac{\dd x}{2\pi}\frac{1}{(x+\ii\epsilon)^2}f(x) = 
\ii f'(0) + \mathcal{P}\left\{\int_{0}^{\infty}\frac{\dd x}{2\pi}\frac{f'(x)}{x}\right\}-\frac{f(0)}{\ii\epsilon},
\end{equation}
where $f(x)$ is a test function, and we already specialised to the case of double poles since this is the only case relevant here. This is true since the Smirnov factorisation for connected and disconnected form factors for an interacting theory still holds. Hence the integrals contain only regularised functions and squared Dirac deltas, up to a divergent term coming from integration over boundaries which is cancelled by the contributions that mix disconnected and connected pieces of the matrix element. In fact it is given by
\begin{equation}
	\label{SemiReg}
	\frac{f(0)}{\ii\epsilon} = 2\int_{0}^{\infty}\frac{\dd x}{2\pi}\frac{f(x)}{x+\ii\epsilon}\delta(x),
\end{equation}
which contains the connected most divergent part of the form factor and the Dirac delta, while $x$ can be considered the difference between two rapidities.

One may wonder whether the outlined procedure is rigorously defined, since we are considering integral expressions instead of discrete sums. However, the procedure amounts to approximating our results up to terms of $\mathcal{O}\left(1/L\right)$, which are sub-leading in the scaling limit.

%%%%%%%%%%%%%%%%%%%%%%%%%%%%%%%%%
\subsection{Proof of (\ref{NDeltaReg})}
\label{Proof 1}
%%%%%%%%%%%%%%%%%%%%%%%%%%%%%%%%%
The main ingredients for this derivation are (\ref{FVFF}) and (\ref{FVStNorm}). The scalar product of two states with by $N$ particle pairs in the finite volume is
\begin{eqnarray}
\label{FVPsStNorm}
&& \langle -I_N,-I_{N-1},\dots,I_N|-J_N,-J_{N-1},\dots,J_N\rangle_L = \delta^2_{I_1,J_1}\dots\delta^2_{I_N,J_N}\\
&&\qquad\qquad = \delta_{I_1,J_1}\dots\delta_{I_N,J_N}=\langle I_1,\dots,I_N|J_1,\dots,J_N\rangle_L ,
\end{eqnarray}
where we used that any power of a Kronecker delta is the Kronecker delta itself and used (\ref{FVStNorm}). Now we recover (\ref{NDeltaReg}) exploiting the fact that the scalar product is nothing but the matrix element of the identity, thus (\ref{NDeltaReg}) follows from (\ref{FVFF}). To get rid of the squared root we have taken advantage of the presence of the Dirac deltas on the left hand side which allows us to consider $\xi_j=\theta_j$ in the function in front of them.

%%%%%%%%%%%%%%%%%%%%%%%%%%%%%%%%%
\section{Linked cluster expansion for Lieb--Liniger model}
\label{Linked Cluster expansion for Lieb-Liniger model}
%%%%%%%%%%%%%%%%%%%%%%%%%%%%%%%%%
In this section we apply the linked cluster expansion directly to the LLM. The required computations are equivalent or even simpler than those performed for the ShGM, thus we are mostly going to present the results obtained, omitting technical details. For the same reason, only the cases $\Psi^\dagger\Psi$ and $(\Psi^\dagger)^2\Psi^2$ are considered, since the generalisations for arbitrary powers follows straightforwardly.

%%%%%%%%%%%%%%%%%%%%%%%%%%%%%%%%%
\subsection{Linked cluster expansion $n=1$}
\label{Linked Cluster expansion 1}
%%%%%%%%%%%%%%%%%%%%%%%%%%%%%%%%%
According to~\cite{IzerginKorepin84}, the form factors of the $\Psi^\dagger\Psi$ operator can be cast in the following form
\begin{equation}
\label{FF1}
F_{2M}^{\Psi^\dagger\Psi}(p_1,\dots,p_M|q_1,\dots,q_M) = \mathcal{C}_M\frac{\sum_{i=1}p_i - \sum_{j=1}q_j}{\prod_{i<j}(p_i-q_j)},
\end{equation}
where $\mathcal{C}_M$ plays the role of a normalisation factor. Given the particular form of the numerator, it is clear why all the form factors vanish when zero momentum eigenstates are taken into account. However, it is less clear to understand how we can get a constant non-zero result. The reason lies in the fact that (\ref{FF1}) still allows non-zero connected form factors in the case of vanishing numerator, as shown by Kormos et al.~\cite{Kormos-09}. The density $n_\mathrm{density}$ is then obtained by re-summing the contributions from different states.

%%%%%%%%%%%%%%%%%%%%%%%%%%%%%%%%%
\subsection{Linked Cluster expansion $n=2$}
\label{Linked Cluster expansion 2}
%%%%%%%%%%%%%%%%%%%%%%%%%%%%%%%%%
From Reference~\cite{KorepinBogoliubovIzergin93} we know that the only non-vanishing form factors are the diagonal ones and that the expectation value over the ground state is zero. Thus we can start from $C^{(\Psi^\dagger)^2\Psi^2}_{22}(t)$, which is given by
\begin{equation}
\label{C22psipsi}
C^{(\Psi^\dagger)^2\Psi^2}_{22}(t) = \frac{1}{2}\int_0^{\infty}\frac{\dd p}{2\pi}\frac{\dd q}{2\pi}\tilde K^*(p)\tilde K(q)\tilde F_4^{(\Psi^\dagger)^2\Psi^2}(-p,p|-q,q)e^{\ii\frac{t}{\mu}(p^2 - q^2)}.
\end{equation}
This term does not contain any divergent part and we do not need to make any further computations if we do not care to extract the stationary value. It can be easily checked that the non-relativistic limit of (\ref{LOFP2}) gives back (\ref{C22psipsi}).

The interesting physics comes in the next order term $C^{(\Psi^\dagger)^2\Psi^2}_{44}(t)$,
\begin{equation}
\label{C44psipsi}
C^{(\Psi^\dagger)^2\Psi^2}_{44}(t) = -\tilde\Gamma t C^{(\Psi^\dagger)^2\Psi^2}_{22}(t) + \tilde Z_2C^{(\Psi^\dagger)^2\Psi^2}_{22}(t) + D'_{44}(t),
\end{equation}
where $\tilde\Gamma$ is given by (\ref{NRmapGamma}) and $\tilde Z_2$ depends on the finite-volume regularisation scheme of the LLM. As in the ShGM, the first term comes from the double pole (spotted also from the linear dependence on time) and the second from the disconnected contribution. The third just collects all the finite valued remainders.

The resummation of the series results in the exponential damping. We can also check that the power-law time dependence arising from the prefactor $C^{(\Psi^\dagger)^2\Psi^2}_{22}(t)$ is correctly predicted to be $t^{-3}$, see (\ref{LateTimeLLn}). 

%%%%%%%%%%%%%%%%%%%%%%%%%%%%%%%%%%%%%%
\medskip

\section*{References}

%\bibliographystyle{../../bibtex/beunsrt}
%\bibliography{../../bibtex/book,../../bibtex/paper,../../bibtex/schuricht}

\end{document}